\documentclass[10pt,conference]{IEEEtran}
\usepackage{cite}
\usepackage{amsmath,amssymb,amsfonts}
\usepackage{graphicx}
\usepackage{textcomp}
\usepackage{xcolor}
\usepackage[hyphens]{url}
\usepackage{fancyhdr}
\usepackage{hyperref}
\usepackage{url}
\usepackage{multirow}
\usepackage{caption}
\usepackage{subcaption}
\usepackage{epsfig}
\usepackage{graphicx}
\usepackage{amsmath}
\usepackage{amssymb}
\usepackage{color}
\usepackage{setspace}
\usepackage{cite}
\usepackage{hhline}   
\usepackage{algorithm}
\usepackage{algpseudocode}
\usepackage{flushend}
\usepackage{url}
\usepackage[normalem]{ulem}
\usepackage{graphics}
\usepackage{tikz}
\usepackage{array}% http://ctan.org/pkg/array
\usepackage{animate}
\usepackage{balance}
\usepackage{enumitem}
\usepackage{booktabs}
\usepackage[skip=2pt,font=footnotesize]{caption}

\newcommand{\hpcayear}{2024}

\newcommand{\hpcasubmissionnumber}{283}
\title{ECO-CHIP: Estimation of Carbon Footprint of Chiplet-based Architectures for Sustainable VLSI}
\def\hpcacameraready{} % Uncomment to build camera-ready version

\newcommand\hpcaauthors{Chetan Choppali Sudarshan$^1$, Nikhil Matkar$^1$, Sarma Vrudhula$^1$, Sachin S. Sapatnekar$^2$, and Vidya A. Chhabria$^1$ }
\newcommand\hpcaaffiliation{$^1$Arizona State University; $^2$University of Minnesota}
\def\aeopen{}           % The artifact is publically available
\def\aereviewed{}     % The artefact has been reviewed
\def\aereproduced{} % The results have been reproduced
\author{
  \ifdefined\hpcacameraready
    \IEEEauthorblockN{\hpcaauthors{}}
      \IEEEauthorblockA{
        \hpcaaffiliation{} \\
      }
  \else
    \IEEEauthorblockN{\normalsize{HPCA \hpcayear{} Submission
      \textbf{\#\hpcasubmissionnumber{}}} \\
      \IEEEauthorblockA{
      }
    }
  \fi 
}

\fancypagestyle{camerareadyfirstpage}{%
  \fancyhead{}
  
  \fancyhead[C]{
    \ifdefined\aeopen
    \parbox[][12mm][t]{13.5cm}{\hpcayear{} IEEE International Symposium on High-Performance Computer Architecture (HPCA)}    
    \else
      \ifdefined\aereviewed
      \parbox[][12mm][t]{13.5cm}{\hpcayear{} IEEE International Symposium on High-Performance Computer Architecture (HPCA)}
      \else
      \ifdefined\aereproduced
      \parbox[][12mm][t]{13.5cm}{\hpcayear{} IEEE International Symposium on High-Performance Computer Architecture (HPCA)}
      \else
      \parbox[][0mm][t]{13.5cm}{\hpcayear{} IEEE International Symposium on High-Performance Computer Architecture (HPCA)}
    \fi 
    \fi 
    \fi 
    \ifdefined\aeopen 
      \includegraphics[width=12mm,height=12mm]{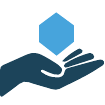}
    \fi 
    \ifdefined\aereviewed
      \includegraphics[width=12mm,height=12mm]{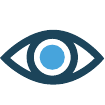}
    \fi 
    \ifdefined\aereproduced
      \includegraphics[width=12mm,height=12mm]{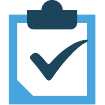}
    \fi
  }
  \fancyfoot[C]{}
}
\begin{document}
\maketitle

%Enables the camera ready header and footer
\ifdefined\hpcacameraready 
  \thispagestyle{camerareadyfirstpage}
  \pagestyle{empty}
\else
  \thispagestyle{plain}
  \pagestyle{plain}
\fi

\newcommand{\hpcaheight}{0mm}
\ifdefined\eaopen
\renewcommand{\hpcaheight}{12mm}
\fi

\vspace{-10mm}
%%%%%%%%%%%%%%%%%%%%%%%%%%%%%%%%%%%%%%%%
%%%%%%%% -- PAPER CONTENT STARTS -- %%%%%%%%%

\begin{abstract}

Decades of progress in energy-efficient and low-power design have successfully reduced the operational carbon footprint in the semiconductor industry. However, this has led to increased embodied emissions, arising from design, manufacturing, and packaging. While existing research has developed tools to analyze embodied carbon for traditional monolithic systems, these tools do not apply to near-mainstream heterogeneous integration (HI) technologies. HI systems offer significant potential for sustainable computing by minimizing carbon emissions through two key strategies: ``reducing" computation by ``reusing" pre-designed chiplet IP blocks and adopting hierarchical approaches to system design. The reuse of chiplets across multiple designs, even spanning multiple generations of ICs, can substantially reduce carbon emissions throughout the lifespan.  This paper introduces ECO-CHIP, a carbon analysis tool designed to assess the potential of HI systems toward sustainable computing by considering scaling, chiplet, and packaging yields, design complexity, and even overheads associated with advanced packaging techniques. Experimental results from ECO-CHIP demonstrate that HI can reduce embodied carbon emissions by up to 30\% compared to traditional monolithic systems. ECO-CHIP is integrated with other chiplet simulators and is applied to chiplet disaggregation considering other metrics such as power, area, and cost. ECO-CHIP suggests that HI can pave the way for sustainable computing practices.
\end{abstract}

\section{Introduction} 
\label{sec:intro}
\noindent
All aspects of computing, from small chips to large datacenters, come with a carbon footprint (CFP) price tag. For several decades, the semiconductor industry has focused on making chips smaller, faster, and less power-hungry, but few efforts have considered the impact on the environment.   The dramatic increase in the demand for compute in the past two decades, fueled by new applications (e.g., artificial intelligence) that demand at-edge and at-cloud-scale computing, has resulted in the information and computing technology (ICT) sector contributing to more than 2\% of the world's CFP~\cite{ict-carbon-report} --  half that of the aviation industry~\cite{commercial-aviation} and projected to surpass it in the next decade if left unchecked. 

\begin{figure}[t]
\centering
\includegraphics[width=0.9\linewidth]{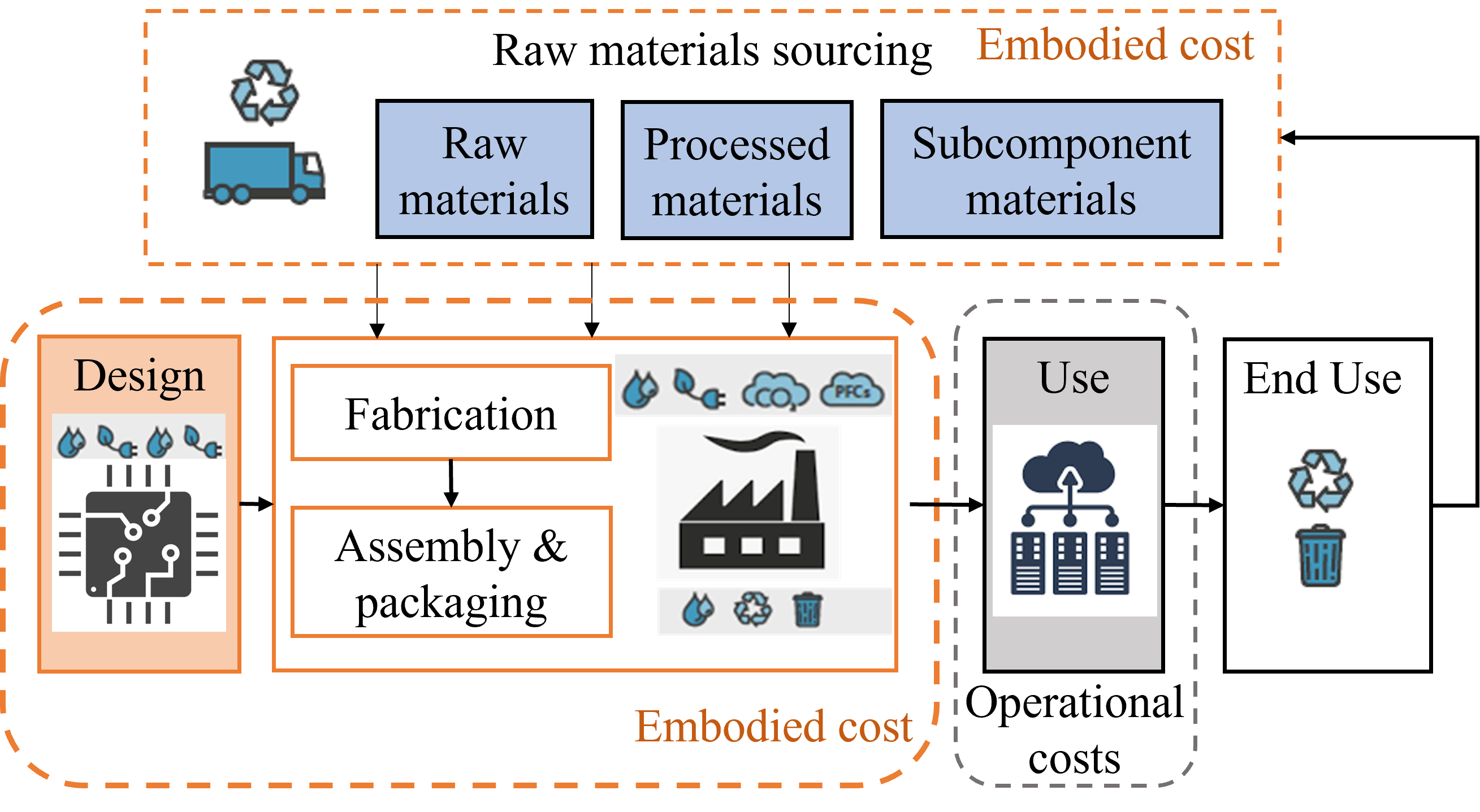}
% \vspace{-4mm}
\caption{Embodied and operational CFP sources in the VLSI supply chain~\cite{imec-wp}.}
\label{fig:lca}
\vspace{-7mm}
\end{figure}

Fig.~\ref{fig:lca} shows the life cycle assessment (LCA) of a semiconductor product and highlights the different sources of greenhouse gases (GHG) in the life of product.  The operational costs refer to the CFP generated by the end user, which in the case of a datacenter are the day-to-day activities that draw energy. The embodied costs are the costs that come from design, manufacturing, packaging, and materials sourcing of the server class computation resources in the datacenter.

While technology scaling and electronic design automation have helped to design energy-efficient VLSI systems with lower operational CFP, the environmental footprint has continued to increase over the past decade and is now dominated by carbon emissions from chip design and manufacturing~\cite{act, imec-wp, imec-dtco, apple-report}, i.e., the embodied CFP, especially for low-power edge devices. 
It is imperative to look beyond the metrics of low power and energy efficiency and include total CFP (embodied and operational) as a first-order optimization metric~\cite{act+, act} for sustainable use of today's modern computing devices.  Several technology companies have pledged to limit their CFP~\cite{apple-report, facebook-report, microsoft-report}, and this can only be achieved by adopting approaches that are cognizant of CFP.

Further, with Moore's law slowing down and SRAM and analog components not scaling~\cite{AMD-Zen2, cmos-scaling-trends},  the way forward towards sustaining Moore's law to the era of trillion-transistor multi-functional systems and beyond is through HI~\cite{subu-iyer, Gelsinger22}. Instead of building system functionality on a single die, HI integrates a set of chiplets, each corresponding to the single die of today, onto a substrate that enables high-density, high-bandwidth chiplet-to-chiplet interconnections. Recent and upcoming advances in HI, including the rapid shrinking of bond pitches between chiplets and interposers~\cite{Mahajan16,Hou17,fevoros,Chen20,HIR21}, enable the design of increasingly sophisticated integrated systems that not only improve cost due to smaller dies and higher yields but also improve power and performance as shown in recent commercial products~\cite{AMD-Zen2}. 

In fact, from a sustainability perspective, heterogeneous chiplet-based systems make a compelling case for CFP evaluation. With the higher yields due to smaller dies, the ability to ``mix and match" chiplets in different technology nodes (older nodes have lower defect densities and lower CFP than advanced nodes), lower silicon wastage on the periphery of wafers due to smaller die sizes, and the savings on design costs due to the availability of pre-designed chiplets, HI systems have the potential to pave a path towards greener VLSI systems. This calls for CFP estimation tools at the architecture level that can model HI \textbf{systems} and not just monolithic dies as in~\cite{act, act+, kaya-ghent}.  New CFP models are needed to account for packaging overheads, silicon fabrics, and multi-die system integration. Such models can be embedded into the emerging HI design methodologies to optimize HI systems for power, performance, area, cost, {\em and carbon}.

Inspired by the principles of environmental sustainability --  ``reduce" and ``reuse" -- in this paper, we evaluate the potential of HI systems towards sustainable computing. HI systems have the potential to lower CFP by ``reducing" carbon emissions by reducing the computation involved in designing each component from scratch and by ``reusing" pre-designed and bulk-manufactured general-purpose chiplet blocks through hierarchical approaches. The ability to reuse chiplets across several designs, not only in the current generation of ICs but even in the next generation, can massively amortize the embodied CFP just as it amortizes the dollar cost~\cite{chiplet-actuary}.

In this paper, we introduce ECO-CHIP, a tool to {\bf E}stimate the {\bf C}arb{\bf O}n footprint of {\bf CHIP}let-based architectures for sustainable VLSI. ECO-CHIP is tailored explicitly for heterogeneous systems that incorporate advanced packaging techniques. Our goal is to demonstrate the potential of such systems in reducing CFP compared to large monolithic dies, even after accounting for the overhead associated with packaging. 

% To estimate the manufacturing CFP of chiplets in heterogeneous systems, we extend existing models from~\cite{act} and~\cite{imec-dtco} with modifications to account for area, yield,  defect densities, and energy-efficiency of process equipment which all scale as the technology matures.

% Our packaging CFP models account for four types of wafer-level packaging architectures: RDL Fan-out, silicon bridges such as embedded multi-die interconnect bridges (EMIB), passive interposers, and active interposers. Further, we consider the overheads concerning inter-die whitespace and additional inter-die routing overheads (network-on-chip protocols) as a part of the packaging CFP estimation.  For design CFP, we employ a simplified model based on design compute time and EDA tool productivity with every technology node to predict the carbon emissions associated with the design phase. 

\noindent
The key contributions of our paper are as follows:
\begin{enumerate}
    \item To the best of our knowledge, this is the {\bf first work to propose heterogenous chiplet-based systems as a direction toward sustainable VLSI.} The paper highlights how heterogenous chiplet-based architectures enable {\bf ``reuse" and ``reduced" chip design and manufacturing despite advanced packaging overheads.}
    \item We develop a novel architectural-level analysis tool, {\bf ECO-CHIP, to estimate the total CFP (design, manufacturing, packaging, and operation) of HI systems}, accounting for various packaging architectures, scaling, yield, process equipment energy efficiency, lifetimes, duty cycles, wafer silicon wastage,  and EDA tool productivity.
    \item This is the {\bf first work to build a CFP estimator for a variety of HI packaging architectures,} considering whitespaces on the package substrate/interposer and inter-die communication overheads.
    \item We evaluate our tool and the potential HI has in reducing CFP on a diverse set of industry testcases (mobile processors, GPUs, and CPUs) and find that {\bf HI systems can reduce embodied CFP by up to 70\%} due to the ability to ``mix and match" technology nodes, yield improvements due to smaller die sizes, and availability of predesigned and bulk-manufactured chiplets. 
    \item We demonstrate the {\bf use of our tool in performing SoC to chiplet disaggregation, considering other metrics such as area, power, and dollar cost} by integrating with other third-party chiplet models. 
\end{enumerate}

\noindent
We open-source ECO-CHIP for broader access and awareness within the community and is currently available at~\cite{github}. 
\section{Motivation and Background}
\label{sec:preliminaries}
\subsection{HI enabling sustainable computing}
\noindent
HI has offered a feasible approach for cost-effective chip design to help sustain Moore's law. A HI system splits a large SoC into multiple smaller dies, referred to as chiplets, where each chiplet may have different functionality, potentially built in different process nodes or designed by separate vendors reducing both design time and cost. All chiplets are integrated into a single package.  HI systems have great potential to lower the embodied CFP associated with design and manufacturing when compared to monolithic systems due to several reasons:

\begin{figure}[tb]
\centering
\includegraphics[width=0.9\linewidth]{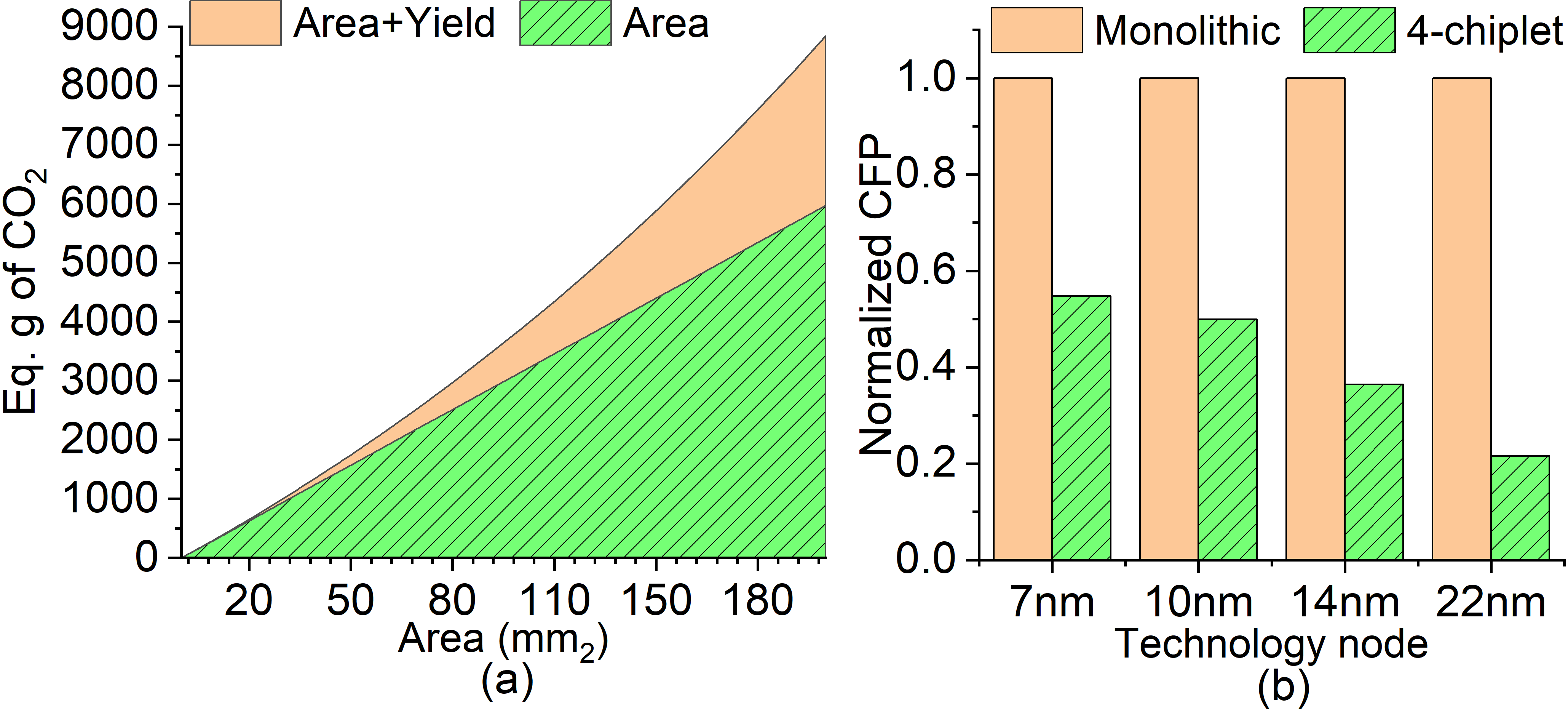}
%\vspace{-2mm}
\caption{(a) Embodied CFP versus area of the chip. (b) Comparison of manufacturing CFP of the large monolithic NVIDIA GA102 GPU and a 4-chiplet-based architecture of the GPU.}
\vspace{-5mm}
\label{fig:monolith-vs-chiplet}
\end{figure}

% \begin{enumerate}[leftmargin = 0.2in]
% \item 
\noindent
\textit{(1)} \underline{\it Yield and area} As we pack more functionality and logic onto the same monolithic IC, the increase in the area increases the CFP due to an increase in materials needed for manufacturing and a decrease in yield. Fig.~\ref{fig:monolith-vs-chiplet}(a) shows a result for an industry testcase in a 10nm technology (using the techniques described in Section~\ref{sec:eco-chip}). We sweep the area of the monolithic SoC up to 200mm\textsuperscript{2} and observe an exponential increase in the associated manufacturing CFP (expressed in equivalent grams of CO$_2$) due to lower yields. In a HI system, each of the smaller dies can be manufactured with a significantly lower environmental cost.  For example, Fig.~\ref{fig:monolith-vs-chiplet}(b) compares the CFP of a monolithic NVIDIA GA102 GPU testcase against a 4-chiplet representation of the GPU where the memory and analog components are on independent chiplets, and the large digital block is split into two smaller chiplets. The CFP of the 4-chiplet design is normalized to the monolithic design's CFP for different technology nodes. The 4-chiplet GPU has significantly lower manufacturing CFP even after including the carbon overheads from advanced packaging due to the higher yields when compared to monolithic systems.

\noindent
\textit{(2)} \underline{\it Technology node ``mix and match"} In a chiplet-based system, dies can be implemented in different technology nodes and integrated into a single package. With analog and SRAM blocks not scaling at the same rate as digital logic, several design houses~\cite{AMD-Zen2, intel-meteor-lake} use chiplets in older technology nodes for memory controllers and analog logic. As pointed out in~\cite{imec-wp}, the CFP to manufacture chips in older technology nodes is much lower than for newer technology nodes due to lower defect densities (better yields), fewer lithography steps due to fewer back-end-of-line (BEOL) or front-end-of-line (FEOL) layers, and the better energy-efficiency of lithography equipment involved in manufacturing older technology nodes with today's latest manufacturing equipment. Typically, even EDA tools scale with technology, and the latest versions of EDA tools can perform design faster with better quality of results on an older technology node~\cite{eda-tool-scaling} due to continuous improvements made by the EDA industry. 

% It is important to note that even though the die area increases due to lower transistor densities in older technology nodes, the benefits from improved yield, matured process, and fewer lithography steps outweigh any increase in CFP due to an increase in area (as we will show in Section~\ref{sec:results}).  

\begin{figure}[tb]
\centering

\includegraphics[width=0.9\linewidth]{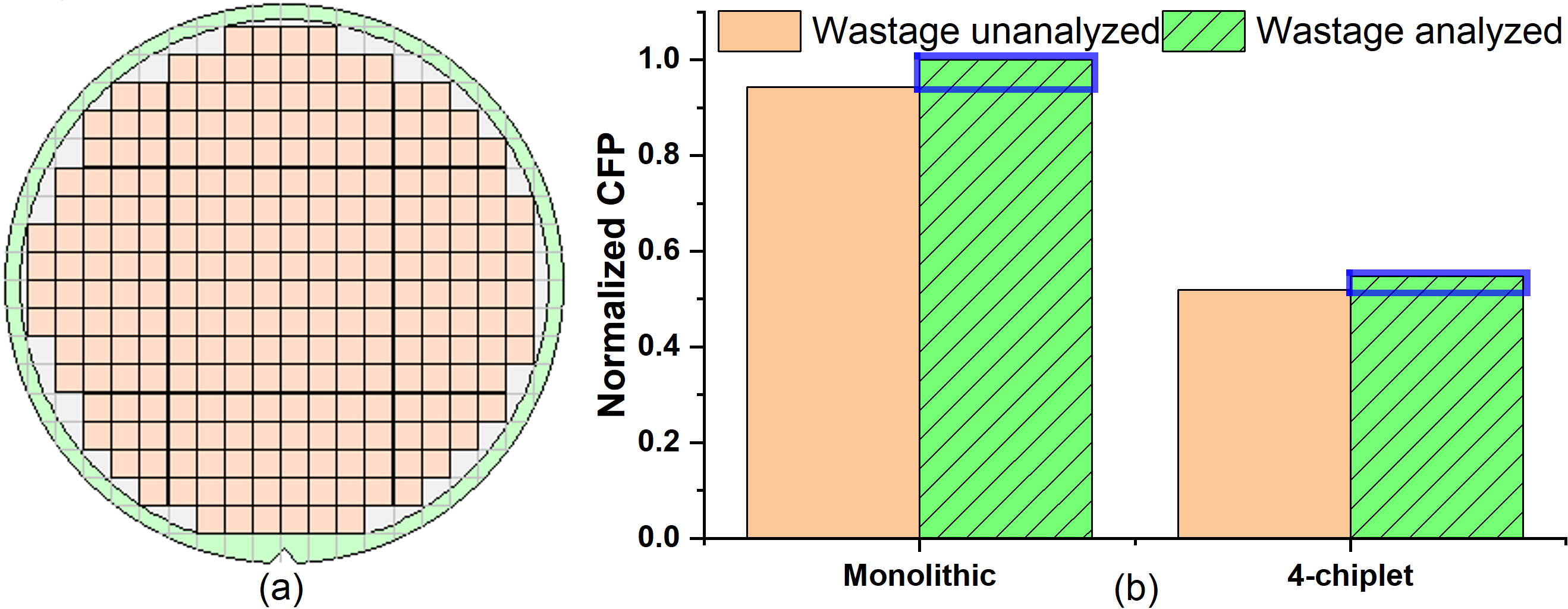}
%\vspace{-2mm}
\caption{(a) Dies on a wafer highlighting the green and white regions of the wafer that are wasted. (b) Comparison of the manufacturing CFP with and without analyzing wastage around wafer periphery for GA102 GPU monolithic and 4-chiplet-based architecture on a 450mm wafer.}
\vspace{-3mm}
\label{fig:area-wasted}
\end{figure}   

% \item 
\noindent
\textit{(3)} \underline{\it Chiplet reuse - Reduced design and manufacturing CFP} Reusing existing silicon-proven die not only saves design time to market directly but also saves the associated design-time CFP. Moreover, composing custom chips out
of small, algorithmic chiplets, reusable across diverse designs, can effectively amortize the non-recurring engineering (NRE) dollar costs and CFP across several different designs~\cite{chopin}. 
A further reduction in manufacturing CFP is due to the reduction in the area wasted around the periphery of the wafer during manufacturing.  Fig.~\ref{fig:area-wasted}(a) shows an image of a wafer with die slices, the green circle around the periphery of the wafer and the white regions are unusable. This area wasted is normalized across the number of dies per wafer (DPW). Smaller-sized dies have lower area wasted compared to larger-sized dies as they can fit a larger number of dies per wafer and also improve the area utilization of the wafer due to the geometric discretization problem. Chiplet-based systems allow a reduction in total number of wafers used due to the fact that more dies can be extracted from a single wafer for smaller die sizes. Fig.~\ref{fig:area-wasted}(b) shows normalized manufacturing CFP of monolithic GA102 testcase and 4-chiplet version of the same testcase with and without considering wastage.

\subsection{HI and packaging architectures}
\noindent
% HI has offered a feasible approach for cost-effective chip design which can help sustain Moore's law. An HI system splits a large SoC into multiple smaller dies, referred to as chiplets, where each chiplet may have a different functionality, potentially built in different process nodes or designed by separate vendors reducing both design time and cost. All chiplets are integrated into a single package.
HI systems are available in different packaging architectures, as shown in Fig.~\ref{fig:packaging-arch}, varying in cost and complexity.   Multi-die HI systems may have anywhere between two to tens of different chiplets. Depending on the number of chiplets, budgeted cost, and complexity, the choice of packaging architecture for heterogeneous systems is different~\cite{lau22}.  We describe four advanced packaging and integration technologies:

\begin{figure}[t]
\centering
\includegraphics[width=0.9\linewidth]{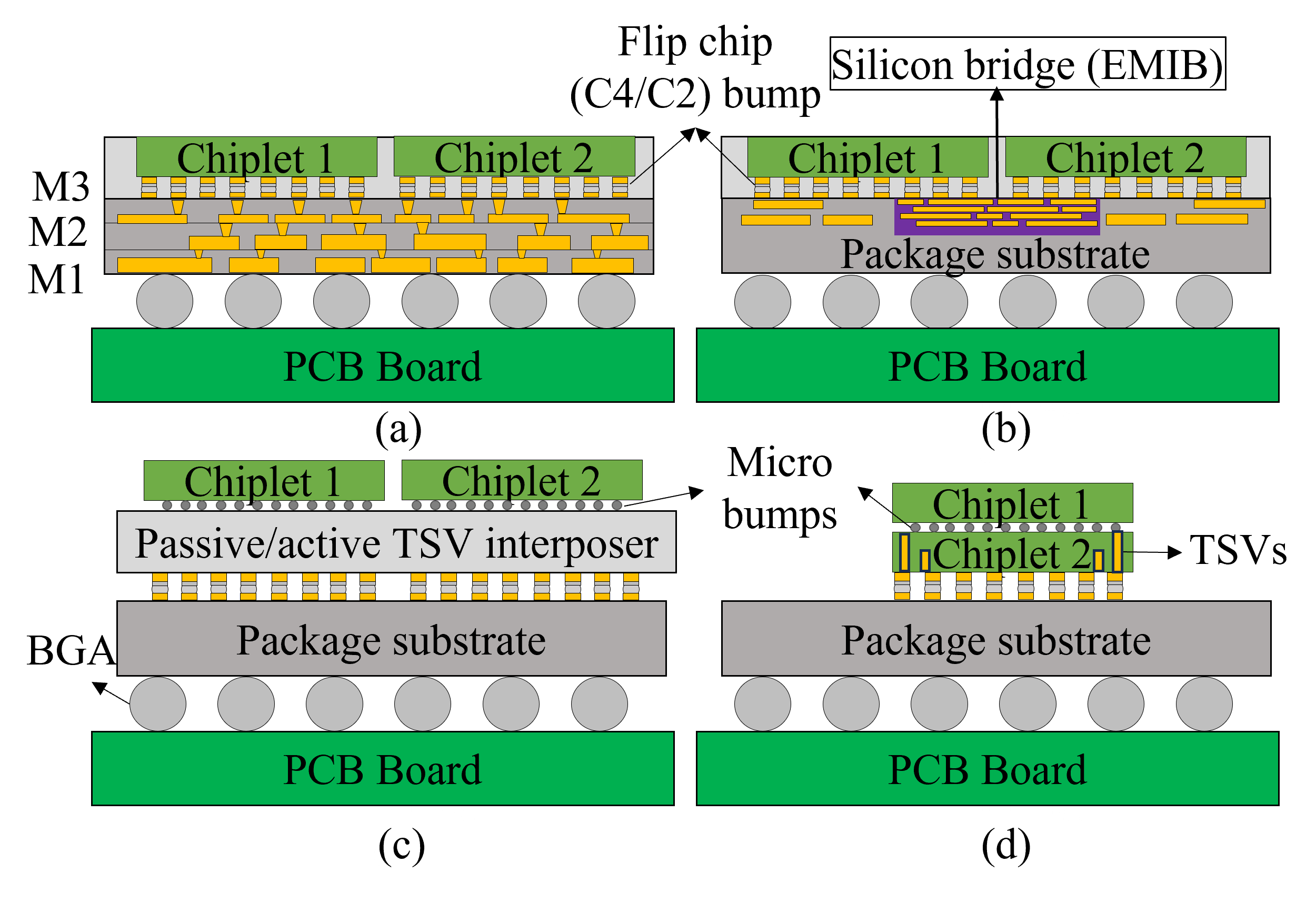}
\vspace{-2mm}
\caption{Packaging architectures: (a) RDL fanout, (b) thin-film and silicon bridge architecture (Intel's EMIB, TSMC's LSI), (c) 2.5D integration with active or passive interposers and (4) 3D stacking with TSVs and microbumps.}
\vspace{-6mm}
\label{fig:packaging-arch}
\end{figure}

\noindent
\textit{(1)} \underline {\it RDL fanout integration} (Fig.~\ref{fig:packaging-arch}(a)) involves the integration of multiple chiplets on a package substrate or fanout redistribution layer (RDL) substrate. Typically the package substrate consists of 3-4 RDL metal layers with linewidths and spacings (L/S) varying from 6/6$\mu$m to 10/10$\mu$m.

\noindent
\textit{(2)} \underline {\it Thin film and silicon bridge-based integration} uses a package substrate that has thin-film layers defined as embedding fine metal RDLs, or a silicon bridge on top of a build-up organic package substrate or in a fanout epoxy molding compound substrate as highlighted in Fig.~\ref{fig:packaging-arch}(b). Intel's embedded multi-die interconnect bridge (EMIB) and TSMC's local silicon interconnect (LSI) are examples of this technology. The package uses local silicon bridges to host ultra-fine L/S structures (2$\mu$m) for die-to-die communications. 

\noindent
\textit{(3)} \underline {\it Passive and active interposer-based integration} involves multiple chiplets in the package that are supported by a through-silicon via (TSV)-less or TSV-based active/passive interposer, and then attached to a package substrate, as shown in Fig.~\ref{fig:packaging-arch}(c). This technology is typically termed a 2.5D architecture. The active interposer consists of both FEOL and BEOL layers, while the passive interposer consists of BEOL layers only, both of which are typically implemented in an older technology node. 

\noindent
\textit{(4)} \underline{\it 3D integration} uses active interposers to support the chiplets, which are then attached to the packaging substrate, or stacks multiple chiplets over the packaging substrate, connected through microbumps or through silicon vias (TSVs), as shown in Fig.~\ref{fig:packaging-arch}(d), or direct bumpless bonding~\cite{fevoros}. With a face-to-back (F2B) stacking of chiplets TSVs are used for connections across tiers; with face-to-face stacking of chips, micobumps are used for inter-die die communication. 

\subsection{Summary}
\noindent
HI has opened up a large new design space previously unexplored by architectural-level carbon simulators~\cite{act, act+, kaya-ghent}. This design space theoretically has significant potential to lower CFP.  However, the exploration of this space requires the development of models that can account for 
the different possible design decisions that impact the CFP.
For instance, the above four described packaging architectures differ in their yields, assembly process, and material used and therefore have different CFPs. EMIB consists of high-density interconnects with fine L/S, typically having lower yields than the larger RDL layers in fanout packaging. Interposer-based integration strategies typically use more materials, and layers, and have a more complex manufacturing process compared to the fanout RDL and EMIB architectures which result in larger CFP. 

Our work, ECO-CHIP, evaluates the CFP for these packaging architectures and the potential HI systems have for sustainable computing. ECO-CHIP is integrated with other third-party chiplet-based tools such as ORION 3.0~\cite{orion} to model inter-die communication overheads accurately and also integrates with cost~\cite{pgupta-cost} and power estimators~\cite{orion} to perform chiplet disaggregation considering power, area, and cost. 
\section{ECO-CHIP framework}
\label{sec:eco-chip}

\begin{figure}[t]
\centering
\includegraphics[width=\linewidth]{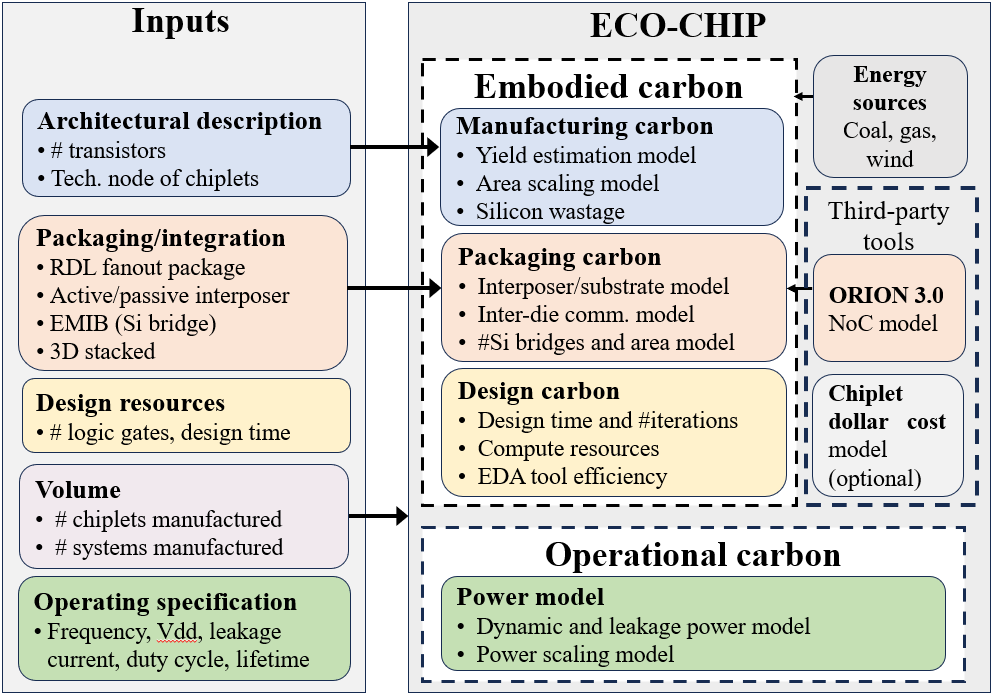}
\caption{ECO-CHIP framework highlighting inputs and the models developed to output embodied and operational carbon. The embodied CFP estimation accounts for the CFP from packaging (red), manufacturing (blue), and design (yellow). The operational CFP (green) is estimated from power models. }
\vspace{-5mm}
\label{fig:eco-chip-top-level}
\end{figure}

\noindent
A high-level description of ECO-CHIP is shown in Fig.~\ref{fig:eco-chip-top-level}. It highlights the inputs and the models developed to generate embodied and operational carbon as output. This section describes the ECO-CHIP framework in detail.

\subsection{ECO-CHIP input description}
\noindent
\textit{(1)} {\underline {\it Architectural description}} High-level description of the SoC or the chiplet-based system, including the predicted number of transistors or logic gates in the chiplet/SoC and the technology node each chiplet is implemented in.

\noindent
\textit{(2)} \underline {\it Choice of packaging integration technique} ECO-CHIP supports four different advanced packaging architectures (described in Section~\ref{sec:preliminaries}). The RDL-fanout-based packaging requires specifying the number of metal layers used and the technology node; the Si bridge-based architectures require specifying the range of the bridge, area of the bridge, and the number of BEOL layers in the bridge; the interposer-based packages require specifying the technology node of the interposer and the number of BEOL layers, and the 3D integration, i.e., chip-to-chip stacking requires specifying the pitch and size of the TSV's and microbumps to estimate the CFP overheads specific to HI.  

% \noindent 
% \redHL{\textit{(3)}  {{\underline {\it Design resources:}}}} The number of design iterations 

\noindent
\textit{(3)} {\underline {\it Operating specification}} To measure operational carbon, we consider the frequency of operation, lifetime, voltage supply, leakage current, and device usage times as input. 
 
\noindent 
\textit{(4)} {\underline {\it Energy sources and chiplet/system volumes}} Other inputs include carbon intensity of different sources and volumes of chiplets and systems manufactured to scale the total CFP based on sources of energy and across the number of parts.

% ECO-CHIP takes an architectural description of an SoC/chiplet-based testcase as input. The architectural description includes the predicted number of logic gates or transistors in the chiplet/SoC, and the technology node of each chiplets. ECO-CHIP uses the information along with

% ECO-CHIP models the area of each chiplets based on the transistor densities of the technology node it is implemented in.  

\subsection{ECO-CHIP total CFP estimation}
\noindent
With the diversity of today's chips, ranging from low-power processors at the edge where embodied CFP dominates operation CFP to high power consuming GPUs in a datacenter, where operational CFP dominates embodied CFP, optimizing the total CFP (embodied and operational) is essential for generalizable sustainable computing. ECO-CHIP models the total CFP ($C_\text{tot}$) as the sum of the operational CFP ($C_\text{op}$) and embodied CFP ($C_\text{emb}$) as:
% \vspace{-2mm}
\begin{equation}
    C_\text{tot}= C_\text{emb} + \text{lifetime} \times C_\text{op}
    \label{eq:tot-cfp}
% \vspace{-1mm}
\end{equation}

\noindent
where $C_\text{emb}$ of the system is the sum of the CFP from the different sources highlighted in Fig.~\ref{fig:lca} and is given by:
\vspace{-2mm}

\begin{equation}
\label{eq:embodied}
    C_\text{emb}= C_{\text{mfg}} + C_{\text{des}} + C_\text{HI} 
    %+ C_\text{mfg,comm} + C_\text{des, comm}
    \vspace{-1mm}
\end{equation}

\noindent
where $C_\text{mfg}$ is the manufacturing CFP of all chiplets, $C_{\text{des}}$ is the design CFP of all chiplets, and $C_{\text{HI}}$ is the overhead from HI including contributions from manufacturing and assembly of the advanced package and any inter-die communication. 

The operational CFP, $C_\text{op}$, is modeled as the carbon intensity of the source of energy during usage ($C_\text{src,use}$) times the energy spent during usage ($E_\text{use}$) and is given by:

\vspace{-3mm}
\begin{equation}
    C_\text{op}= C_\text{src, use}  \times E_\text{use}
    \label{eq:operational}
\end{equation}

In the rest of this section, we detail the models for each of the components in Eqs.~\eqref{eq:tot-cfp},~\eqref{eq:embodied}, and~\eqref{eq:operational}.

\subsection{ECO-CHIP manufacturing CFP estimation}
\noindent
We model the manufacturing CFP of the heterogeneous system as the sum of the manufacturing CFP of each chiplet, $i$, and is given by    $ C_\text{mfg}= \textstyle \sum_{i=1}^{N_C}C_{\text{mfg},i} $. To estimate the manufacturing CFP of each chiplet, we make three essential modifications to~\cite{act, act+} to support the estimation of embodied CFP and perform disaggregation as described below: 

\noindent
\textit{(1)} \underline{{\it Area scaling models}} Since a system disaggregation algorithm or a heterogeneous system requires selecting a technology node for each chiplet, our carbon estimation tool uses transistor density scaling trends from~\cite{tsmc-transistor-density, tsmc-sram-scaling} and transistor counts from our testcase architectures to determine the area of a chiplet in a specific technology node.  The area scaling models are critical to the estimation of CFP as larger chiplet areas in older technology nodes can have larger CFP even though they have lower CFP per unit area (CFPA). We use three different area scaling models for logic, memory, and analog blocks, as each has different transistor densities and, therefore different areas with every technology node.  We evaluate the area of the die as $ A_{die}(d, p) = D_{T}(d, p) \times N_{T}$, where $D_T(d, p)$ is the transistor density for design type $d$ and process $p$, $N_T$ is the number of transistors in the die, and $A_{die}(d,p)$ is the area of die of type $d$ in process node $p$.

\noindent
\textit{(2)} \underline{{\it Yield models}} One of the primary advantages of HI is the cost savings that come with larger manufacturing yields due to smaller die sizes. The increase in yield  compared to a large monolithic die also helps lower CFP. However, if the die is in an older technology node, then $A_\text{die}(d,p)$ must be accounted for as an increase in the area may lower yields which also lowers the CFP, as shown in Fig.~\ref{fig:monolith-vs-chiplet}(a). To estimate the impact of the area on yield and CFP, we use a negative binomial yield distribution model given by~\cite{yield-models, supply-chain, active-vs-passive-2}:

\begin{equation}
\label{eq:yield}
    Y(d, p) = \left ( 1 + \frac{A_\text{die}(d, p) \times D_0(p)}{\alpha} \right ) ^ {-\alpha} 
\end{equation}

\noindent 
where $Y(d, p)$ is the yield of die with area $A_{die}(d, p)$, $D_0(p)$ is the defect density for process $p$, $\alpha$ is a clustering parameter.  It is important to note that the defect density depends on $p$ and scales with technology. This dependence is important to capture because legacy nodes have lower defect densities which result in larger yields, but older technology nodes result in larger $A_{die}$ values leading to lower yields. ECO-CHIP considers these tradeoffs while estimating CFP. 

\noindent
\textit{(3)} \underline{{\it Energy-efficiency of process equipment}} With advances in process equipment, the energy efficiency of the photolithography equipment improves at every step, especially for the more mature technology nodes. We incorporate the energy efficiency of the equipment as a derate factor ($\eta_{eq}$) from~\cite{TSMC-report}. The $C_{\text{mfg},i}$ on a per chiplet basis is given by the sum of the product of the carbon footprint per unit area (CFPA) of manufacturing a die and the area of the die and the product of the CFPA of silicon (CFPA$_\text{Si}$) and amortized area of silicon wafer wasted ($A_\text{wasted}$):

\vspace{-3mm}
\begin{align}
    C_{\text{mfg},i} & = \text{CFPA} \times A_\text{die}(d,p) + \text{CFPA}_\text{Si} \times A_\text{wasted}\\
    \text{CFPA} & = \frac{(\eta_{eq} \times C_{\text{mfg, src}} \times \text{EPA}(p)  + C_{\text{gas}} + C_{\text{material}})}{Y(d, p)}  
    \label{eq:CFPA}
\end{align}

\noindent
where $C_\text{mfg, src}$ is carbon intensity which depends on the energy source of the fab (i.e., renewables vs. non-renewables), which converts the energy consumed into carbon emission. $\text{EPA}$ is the energy consumed per unit area during manufacturing of process $p$ and derived from~\cite{imec-dtco}, $C_\text{gas}$ is the CFP from the greenhouse gas emissions, and $C_{\text{material}}$ is the carbon footprint of sourcing the materials for fabricating the chip/chiplet.

%\blueHL{The wasted silicon area in a wafer with $A_\text{wafer}$ is the area highlighted in Fig.~\ref{fig:area_wasted} in the wafer, which is amortized across then number of manufacturable dies on the wafer and is given by:}
The wasted silicon area in a wafer of area $A_\text{wafer}$ is  highlighted in Fig.~\ref{fig:area-wasted}(a) by the green and white regions. The die cannot occupy zones within its half die diagonal, reducing the usable diameter by $L_\text{d}/\sqrt{2}$. Therefore, the number of dies per wafer (DPW) and the amortized area wasted per die ($A_\text{wasted}$):

\begin{equation}
\label{eq:dpw}
    \text{DPW} =  \left \lfloor {\frac{\pi( \frac{D_\text{wafer}}{2} - \frac{L_d}{\sqrt{2}}) ^{2}}{A_\text{die} }} \right \rfloor 
\end{equation}

\begin{equation}
    A_\text{wasted} = \frac{A_\text{wafer} - (\text{DPW}  \times A_\text{die})}{{\text{DPW}}  }
\end{equation}

%\blueHL{Here its important to note that smaller dies have lesser $A_\text{wasted}$ than larger dies as its a geometric discretization problem which lowers the manufacturing CFP for smaller individually manufactured chiplet systems. It also allows for a larger number of dies per wafer enabling a larger amortization of the wasted area. }
\noindent
where $D_\text{wafer}$ is diameter of the wafer, $L_d$ is the side length of the die, and $A_\text{die}$ is the die area. An important observation here is that smaller dies have lesser $A_\text{wasted}$ compared to larger dies as we can cramp in more DPW. This allows for a larger amortization of the wasted.
% Considering a design with 100mm\textsuperscript{2} area, if this is broken into four chiplets, assuming a manufacturing volume of 100,000 and wafer diameter of 300mm, could reduce 14,778 Kgs of CO$_2$ just by considering the need for fewer wafers.}

\subsection{ECO-CHIP HI-oriented CFP overheads}
\noindent
With the widespread adoption of HI systems, the cost of packaging is projected to dominate design~\cite{mdpi}.  Although there are several sustainability reports from large semiconductor manufacturing and design companies, these reports do not specifically break down the contributions from packaging.  The prior art in this area has been limited to wire bond packages and flip chip packages~\cite{lca-package}.  Since HI has opened up a previously unexplored design space, it requires developing models that can account for the different possible design decisions in the HI system that impact the CFP. In particular, decisions related to the choice of the package ($C_\text{package}$), whitespace on the package substrate or interposer ($C_\text{whitespace}$), and inter-die communication ($C_\text{mfg, comm}$). In our work, we measure the CFP from these three sources as described below:

\noindent
\textit{(1)} \underline{{\it Package-related overheads} ($C_\text{package}$)} We develop models for the four different packaging architectures, described in Section~\ref{sec:preliminaries} based on architectural descriptions, materials, and packaging technology nodes from~\cite{lau22}, CFP estimates from~\cite{imec-dtco}, and packaging industry reports~\cite{spil-packaging, amkor-report, ace-report}. 

\noindent
{\it (a) RDL Fanout:}  This packaging architecture uses an epoxy molding compound (EMC) substrate with RDL metal layers patterned to make connections between the chiplets as shown in Fig.~\ref{fig:packaging-arch}(a). Our CFP model uses the energy per unit area per metal layer (EPLA) from a manufacturing fab to determine CFP overheads with the RDL layers. Based on the number of layers, the yield of the layers, and EPLA, we determine the embodied CFP of an RDL package as:

\begin{equation}
\label{eq:rdl-cfp}
    C_\text{RDL} = \frac{L_\text{RDL} \times \text{EPLA}_\text{RDL}(p) \times C_\text{pkg, src} \times A_\text{package}}{Y(\text{RDL}, p)} 
\end{equation}

\noindent
where $\text{EPLA}_\text{RDL}(p)$ is the energy consumed in patterning a single RDL layer in process $p$ per unit area, $C_\text{pkg,src}$ is the carbon intensity of the packaging fab which is based on the source of energy (renewable or non-renewable sources), $L_\text{RDL}$ is the number of layers of RDL in the package substrate, $Y(\text{RDL}, p)$  is the yield of the RDL in process $p$ estimated using Equation~\eqref{eq:yield}, and $A_\text{package}$. The area of the package substrate is estimated after considering the whitespace and routing overheads and described later in this section. 

\noindent
{\it (b) Silicon bridge:} A silicon bridge is a high-density interconnect between two chiplets, and we model its CFP similar to the CFP of the RDL fanout-based package except that they have lower linewidth and spacing (L/S) and, therefore, lower yields when compared to RDL fanout. Our model uses the EPLA values from~\cite{imec-dtco} for an advanced technology node lower metal layer with ultra-fine L/S. These high-density interconnects do not span the entire area of the package substrate but are local to a region in the package based on the floorplan of the chiplets. The number of silicon bridges and their placement depends on the chiplet floorplan and bandwidth requirements. In our work, we consider bridge ranges and typical bridge areas from Intel's EMIB silicon bridge specification~\cite{emib} as input to determine the number of bridges that must be used. An additional bridge is added if the two adjacent dies that must be connected through silicon bridges have overlapping die edges larger than the range. The CFP of a silicon bridge-based packaging architecture is given by:

\vspace{-3mm}
\begin{equation}
    C_\text{bridge} = \frac{N_\text{bridge} \times L_ \text{bridge} \times \text{EPLA}_\text{bridge}(p) \times C_\text{pkg, src} \times A_\text{bridge}}{Y(\text{bridge}, p)} 
    \label{eq:emib-cfp}
\end{equation}

\noindent
where $L_\text{bridge}$ is the number of metal layers in the bridge, $A_\text{bridge}$ is the area occupied by the silicon bridge in the package substrate, $N_\text{bridge}$ is the number of silicon bridges, $Y(\text{bridge}, p)$ is the yield of fabricating the silicon bridge in process $p$ in the bridge cavity, $\text{EPLA}_\text{bridge}(p)$ is the energy per unit layer per unit area of patterning the silicon bridge in process $p$.

\noindent
{\it (c) Active interposer:} Active interposers are manufactured to include transistor devices within the interposer, providing several unique capabilities not possible with passive interposers.  We model these interposers as an additional large die that is typically in an older technology node compared to the chiplets. However, unlike a regular chiplet, the active region is only restricted to local regions with routers and repeaters. 

We use a similar model based on Eq.~\eqref{eq:CFPA} to estimate CFP overhead from active interposer. Interposer-based packaging architectures have higher CFP when compared to the RDL fanout-based packaging and EMIB-based packaging as the interposer acts as an additional large silicon die that spans the entire area of all the chiplets put together with BEOL layers across the entire interposer and active FEOL layers locally in those areas that have routers or repeaters. 

\noindent
{\it (d) Passive interposer:} Unlike active interposers, passive interposers only contain metal interconnect, so they cannot include active logic like routers, or repeaters in the interposer. We model the CFP of the passive interposer in a similar way as Equation~\eqref{eq:rdl-cfp} on a per unit area and per layer basis.

% It is important to note here that the overheads from the package substrate are the same across all architectures and will also be a part of the monolithic system. Therefore, we do not account for the CFP due to package substrates.  The package substrate of a HI system is larger than that of a monolithic system, and the additional CFP due to the area overheads are considered after estimating the packaging area using our whitespace estimator algorithm.  

% \begin{figure}[t]
% \centering
% \includegraphics[width=\linewidth]{figs/active-vs-passive.png}
% \vspace{-8mm}
% \caption{Router location in (a) active and (b) passive interposer.}
% \vspace{-6mm}
% \label{fig:active-vs-passive}
% \end{figure}

\noindent
{\it (e) 3D integration:}
This packaging architecture stack chiplets as shown in Fig.~\ref{fig:packaging-arch}(d) to minimize the 2D footprint and maximizes bandwidth where inter-chiplet communication is performed with TSVs, microbumps, or hybrid bonds. Our CFP model uses the energy per unit area per metal layer (EPLA) from a manufacturing fab to determine CFP overheads with the TSVs, microbumps, and hybrid bonds~\cite{act+, imec-dtco}. The number of TSVs, bumps, or bonds depends on the size of the chip and its pitches. 
In our work, we assume a dense network of TSVs, bumps, or bonds placed at the minimum pitch~\cite{HIR21} to maximize inter-chiplet bandwidth. We use the TSV diameter and pitch values~\cite{HIR21, tsv-pitch}, hybrid and microbump pitch values from~\cite{liang22} and EPLA to determine the embodied CFP as: 
% \vspace{-4mm}
\begin{equation}
    C_\text{3D} = \frac{N_\text{TSV, bump, bond} \times \text{EPA}_\text{TSV, bump, bond}(p) \times C_\text{pkg, src}}{Y(\text{3D}, p)} 
\end{equation}

\noindent
where $N_\text{TSV, bump, bond}$ is the number of TSVs/bumps/bonds, $Y(\text{3D}, p)$ is the yield of the 3D package assembly accounting for misalignments of bumps the bonding yield between chiplets. $\text{EPA}_\text{TSV, bump, bond}(p)$ is the energy per unit area of patterning the TSV or manufacturing the bump in process $p$.

\noindent
\textit{(2)} \underline{{\it Inter-die communication overheads} ($C_\text{mfg, comm}$)} Unlike EMIB and RDL-based packaging architectures, which are limited to supporting few (four - five) chiplets~\cite{lau22}  interposer based 2.5D architectures support tens of chiplets while 3D integration techniques can have 2-3 tiers of logic. However, both interposer-based techniques and 3D integration techniques come with large inter-die communication overheads which require are protocols such as network-on-chip (NoC). To support an NoC router, each chiplet must be equipped with a network interface controller (NIC).  In passive interposers, router modules must be placed within the chiplets, contributing to chiplet area and degrading yield and $C_{\text{mfg},i}$ while with active interposers, router modules can be moved from the chiplets to the interposer, reducing the area in the chiplets and therefore improving chiplet yield and  $C_{\text{mfg},i}$ compared to passive interposers.

To estimate the CFP overheads of routing, we use a third-party tool, ORION 3.0~\cite{orion} and~\cite{active-passive-interposers}. ORION is used to estimate the power overhead due to the additional inter-die communication NoC circuitry and~\cite{active-passive-interposers} is used to estimate the area overhead. The work~\cite{active-passive-interposers} models the network on interposers (NoI/NoC) area by including flit width, bidirectional port counts, and microbump pitches. However, it provides area values for only a small set of specific technology nodes (11nm to 65nm), we scale the values for the technology nodes we consider. The NoC area is added to either the chiplets or the interposer based on active or passive interposer and implemented in the same technology node as that of the chiplet or the interposer.  Although ORION 3.0, supports 45nm and 65nm technologies, we modify the parameter files for the appropriate technology node.  ORION 3.0 models the power of the NoC by estimating the number of instances based on the microarchitectural parameters, including the number of ports, flit width, and buffers~\cite{orion, active-passive-interposers, dsent}. The CFP overhead for interposer-based NoC routers for inter-die communication is given by: $C_\text{mfg, comm} = \text{CFPA} \times A_\text{router}(d, p)$, where $\text{CFPA}$ is defined in~\eqref{eq:CFPA}. For the passive interposer, $A_\text{router}(d,p)$ is added to the area of the chiplet, after which yield and $C_\text{mfg, i}$ is calculated. For active interposers, the carbon contribution of $A_\text{router}(d, p)$ is used to add to the embodied CFP of the system. 

It's important to note that for passive interposers, the NoC is implemented in the same technology node as the chiplet, which is a more advanced node than those routers that are a part of the package. Therefore, routers for passive interposers are of lower areas than the active interposer router in an older technology node.  For EMIB- and RDL-based packages, there are additional communication overheads for PHY~\cite{emib} interfaces that are typically part of the chiplet itself. These interfaces are typically designed as IPs and have small additional areas when compared to the chiplets.

\noindent
\textit{(3)} \underline{{\it Whitespace overheads} ($C_\text{whitespace}$)} To estimate the area of the package substrate or interposer, ECO-CHIP uses a whitespace or system area estimation algorithm. The algorithm performs recursive bi-partitioning to build a slicing floorplan~\cite{slicing-floorplan} of the chiplets on the package substrate/interposer. An initial two-way partition is created by assigning the chiplets (sorted in decreasing order of their area), one by one, to the partition with a lesser total weight. Our model uses the area of each partition as the weight, which results in an area-balanced initial partition.
The bi-partitioning procedure is then used recursively within each partition to perform a K-way partition of the chiplets by first creating two equal-sized partitions, then independently dividing each of these into two subpartitions each, and so on till a partition contains only one chiplet. This effectively creates a full binary tree where each leaf node is a chiplet and each internal node represents a partition. The overall floorplan and its area can be derived by processing the partitions and chiplets within the tree.

For each leaf node, processing involves setting the orientation and aspect ratio of the chiplet to get a bounding box. At the internal nodes, this involved combining two subpartitions together, accounting for whitespace overheads. There are two sources of whitespace overheads: (i) spacing between two subpartitions due to chiplet spacing constraints~\cite{chiplet_spacing, active-passive-interposers}, (ii) if the two subpartitions are imbalanced in terms of their dimensions, we create a bounding box of the two partitions which will result in additional whitespace. 
The recursive bipartisan floorplan also provides us with interfaces between each pair of chiplets to identify locations for routers, and silicon bridges on the package substrate/interposer.

\subsection{ECO-CHIP design CFP estimation}
\noindent
Although design CFP ($C_\text{des}$) is amortized across the number of chiplets of type $i$ manufactured ($N_{M_i}$) and systems manufactured ($N_S$),  several cutting-edge accelerators, GPUs, and server CPUs are not manufactured in sufficiently large numbers to amortize the cost of design across the number of parts manufactured. We model the design CFP, $C_\text{des}$ as:

\vspace{-2mm}
\begin{equation}
    C_\text{des}    = \sum_{i=1}^{N_C}\frac{C_{\text{des},i}}{N_{M_i}} + \frac{C_\text{des,comm}}{N_S} 
    \label{eq:des-cfp}
\end{equation}

% \vspace{-3mm}
% \begin{equation}
%     C_\text{des}    = \frac{\sum_{i=1}^{N_C}C_{\text{des},i} + C_\text{des,comm}}{N_P} 
% \end{equation}

\noindent
where $C_{\text{des},i} =  t_{\text{des},i} \times P_\text{des} \times C_{\text{des,src}}$ is the design CFP of a single chiplet, $C_\text{des,comm}$ is the design CFP of routers for inter-die communication, $t_\text{{des},i}$ is the CPU compute time it takes to design a chip/chiplet, $P_\text{des}$ is the power consumed by the compute resources (CPUs) used to design the chips, $C_{\text{des,src}}$ is the CFP of the energy source. We model $t_\text{des, i}$ as:

% \vspace{-3mm}
\begin{equation}
    t_{\text{des,i}} =  \frac{t_{\text{verif},i} + (t_\text{{SP\&R},i} + t_{\text{analyze},i}) \times N_\text{des}}{\eta_\text{EDA}}
    \label{eq:des-time}
\end{equation}

\noindent
where $t_{\text{verif},i}$, $t_{\text{SP\&R},i}$, $t_{\text{analyze},i}$ are the compute time for verification, and a single synthesis, place, and route (SP\&R) run and a single simulation of all analysis respectively, and $N_\text{des}$ is the number of design iterations.   Further, to model the EDA tool improvement with new version releases, we create a near-linear regression model based on productivity for different technology nodes~\cite{eda-tool-scaling} and scale the value of $t_{\text{des,i}}$ by $\eta_\text{EDA}$.  

%We find that there is nearly a 20\% improvement in design time with every previous process node. In addition, we model $t_{\text{verif},i}$ as twice times that of $t_\text{{SP\&R}},i$ due to verification dominating design time to market. 

\subsection{ECO-CHIP operational CFP estimation}
\noindent
We estimate the operational CFP ($C_\text{op}$) by modeling for the total energy of the system during usage $E_\text{use}$ based on the operational specification give by:

\vspace{-3mm}
\begin{equation}
    E_\text{use}  = T_{ON} \times (V_{dd} I_\text{leak} + \alpha C V_{dd}^2 f) 
\end{equation}

\noindent
where $T_\text{ON}$ is the time for which the system is ON, $V_{dd}$ is the supply voltage, $f$ is the average use case frequency of operation of the system (since most systems are not operating at maximum frequency throughout their use), $\alpha$ is average switching activity, $C$ is the load capacitance, and $I_\text{leak}$ is the leakage current of the system. We then substitute $E_\text{use}$ in Eq.~\eqref{eq:operational} to estimate the operational CFP. $E_\text{use}$ also includes the HI-related power overheads, such as inter-die communication modules (NoCs). For battery-operated devices such as a mobile processor, we can directly estimate $E_\text{use}$, based on the battery rating and frequency of recharging~\cite{act, act+}. 
\vspace{-1mm}
% \section{Evaluation of CFP of real-world testcases}
% \label{sec:results}

\section{ECO-CHIP setup and real-world testcases}
\label{sec:setup}
\begin{figure}[tb]
\centering
\includegraphics[width=0.9\linewidth]{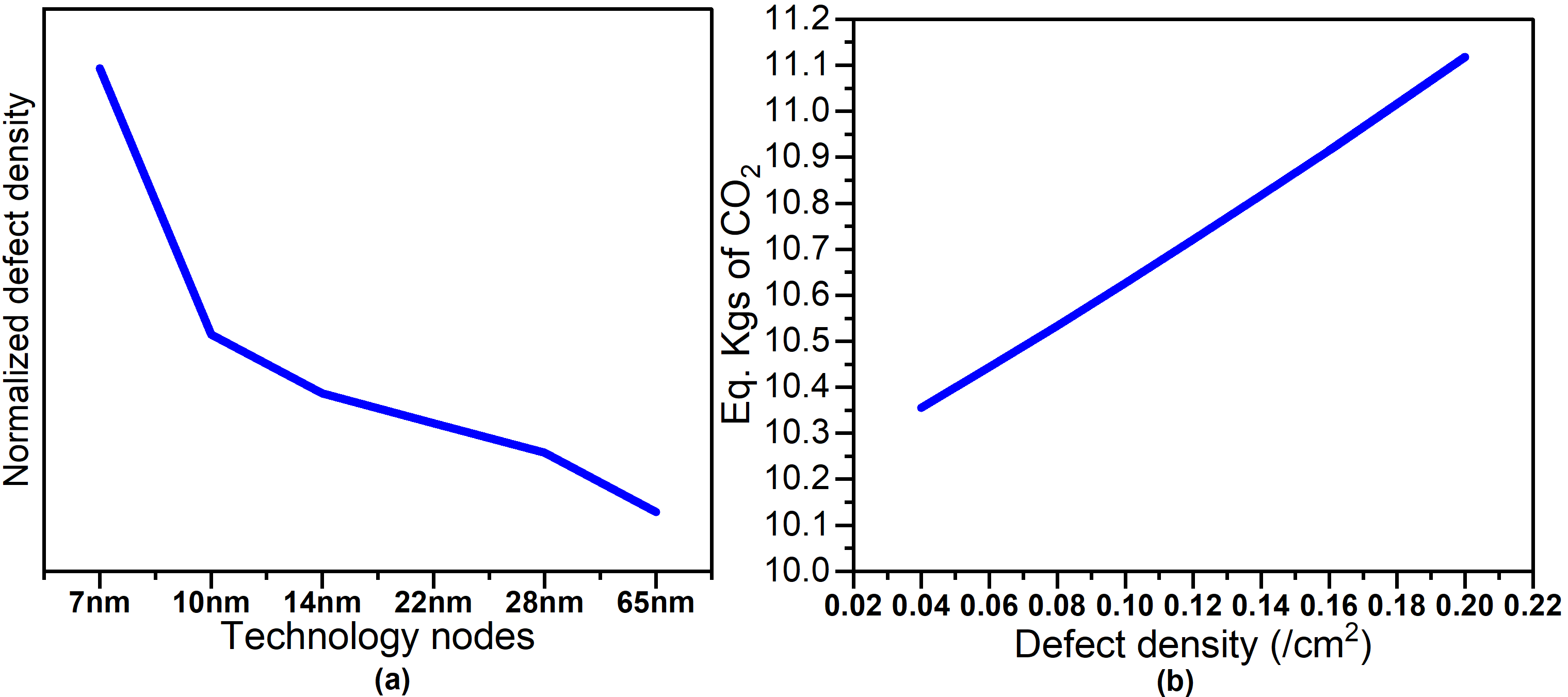}
%\vspace{-2mm}
\caption{(a) Normalized defect density with technology node scaling. Older nodes have lower defect densities~\cite{supply-chain, active-vs-passive-2}. (b) Impact of defect density on total CFP.}
\vspace{-6mm}
\label{fig:def-den-vs-nodes}
\end{figure}

\noindent
\textit{(1)} \underline{{\it Input parameters}} ECO-CHIP uses several input parameters which are listed in Table I with their supported range of values and their sources. For instance, based on the source of energy, whether it is coal, gas, wind, etc., the $C_\text{mfg, src}$ can be a value between 30g -- 700g of CO$_2$ or based on the technology node, the defect densities can be between 0.07 -- 0.3/cm$^2$\cite{supply-chain, active-vs-passive-2}. Defect density varies across nodes, and they tend to mature over time. Fig.~\ref{fig:def-den-vs-nodes}(a) shows the normalized defect density trend over different technology nodes. For a given technology node, defect densities begin to plateau as the node matures with time\cite{anandtech-def-den-tsmc5nm,def-den-variation1,variation2-def-density}. Fig.~\ref{fig:def-den-vs-nodes}(b) shows the variation in total CFP as a function of defect density. Although our simulator can handle a range of technology nodes for packaging and a range of derate factors for $C_\text{mfg, src}$, our results in this section are shown for specific values, i.e., we assume all packaging technology (RDL, EMIB, and active/passive interposers) to be in 65nm, the $D_\text{wafer}$ to be 450mm, and the energy source is from coal at 700g of CO$_2$ per KWh. Based on the testcase, we vary the technology node for each of the chiplets to explore the possible design space and estimate $C_{\text{mfg},i} $. Based on the technology each chiplet is implemented in,  we choose the appropriate values from the specified ranges. 

\begin{table}[tb]
\centering
\label{tbl:knobs}
\caption{Input parameters to ECO-CHIP and their range of values.}
\resizebox{0.9\linewidth}{!}{%
\begin{tabular}{c|crll}
\hline
Model   &     Parameter     &     Value&Unit     &     Source   \\ \hline
 %%%%% section mfg %%%%
 \multirow{8}{*}{$C_{\text{mfg},i}$ }     
 & $D0(p)$ &  0.07 -- 0.3&/cm$^2$  & \cite{supply-chain, active-vs-passive-2} \\
 & $\alpha$ &  3&     &    \cite{supply-chain, active-vs-passive-2}    \\
 & $D_T(d, p)$ &  5 -- 150&MTr/mm$^2$    &\cite{tsmc-transistor-density, tsmc-sram-scaling} \\ 
 & $\eta_{eq}(p)$ &  0 -- 1&   &  \cite{TSMC-report}  \\ %\hline
 & $C_\text{mfg, src}$ & 30 -- 700&g CO$_2$/kWh & \cite{act, imec-dtco}  \\ %\hline
 & $\text{EPA}(p)$ &   0.8 -- 3.5 &kWh/cm$^2$ &    \cite{act, imec-dtco} \\ %\hline
 & $C_\text{gas}$ & 0.1 -- 0.5&kg CO$_2$/cm$^2$ &    \cite{act, imec-dtco}\\% \hline
 & $C_\text{material}$ &   0.5& kg CO$_2$/cm$^2$ &    \cite{act, imec-dtco}\\ 
 & $D_\text{wafer}$  &   25-450& mm  &  \cite{wafer-sizes}\\ 
 \hline

 %%%% pacakge %%%%%
 \multirow{8}{*}{$C_\text{package}$}      
 & RDL tech. & 22nm - 65nm&   &  \cite{lau22, emib, active-passive-interposers} \\ %\hline
 & $\text{EPLA}_\text{RDL}(p)$ & 0.05 -- 0.2 &kWh/cm$^2$   & \cite{act, imec-dtco} \\% \hline
 & $C_\text{pkg, src}$ & 30 -- 700&g CO$_2$/kWh   &  \cite{act, imec-dtco} \\% \hline
 & $L_{RDL}$ & 3 -- 9 & & \cite{lau22} \\% \hline
 & $L_\text{bridge}$ & 3 -- 4& & \cite{emib} \\% \hline
 & Bridge tech. &  22nm -- 65nm&    &  \cite{emib} \\% \hline
 & $\text{EPLA}_\text{bridge}(p)$ &   0.1 -- 0.35& kWh/cm$^2$ & \cite{act, imec-dtco}  \\% \hline
 & Bridge range & 2 $\times$ 2&mm$^2$ & \cite{emib} \\ 
 &  TSV pitch           &   10 -- 45  &  $\mu$m &  \cite{HIR21, tsv-pitch}   \\
 % &  TSV diameter         &   5 -- 10 & $\mu$m &  \cite{HIR21, tsv-pitch}   \\
  &  Microbump pitch         &   10 -- 45 & $\mu$m &  \cite{HIR21}   \\
&  Hybrid bond pitch        &   1 -- 10 & $\mu$m &  \cite{liang22}   \\
 \hline
 
 %%%% communication %%%%%
 \multirow{2}{*}{$C_\text{mfg,comm}$} 
 & Interposer tech. & 22nm -- 65nm&    & \cite{active-passive-interposers}   \\% \hline
 & NoC flit width.  & 512 bits&    & \cite{active-passive-interposers} \\ 
 \hline
 
 %%%% Whitespacae %%%%%
 $C_\text{whitespace}$      &     Chiplet spacing      &     0.1 -- 1 & mm   &  \cite{active-passive-interposers, chiplet_spacing}  \\ 
 \hline
 
 %%%% Whitespacae %%%%%
 \multirow{4}{*}{$C_\text{des}$}    
 &  $\eta_\text{EDA}$ & 0 -- 1&   &  \cite{eda-tool-scaling} \\% \hline
 &  $P_\text{des}$     &   10&W     & \cite{power-per-core} \\% \hline
 &  $N_\text{des}$     &     100&   & \cite{abk-schedule} \\ % \hline
 & $C_\text{des, src}$ & 30 -- 700   & g CO$_2$/kWh   &  \cite{act, imec-dtco}  \\
 \hline

 %%%% operational %%%%
 \multirow{4}{*}{$C_\text{operational}$}    
 &  $V_{dd}$     &     0.7 -- 1.8 &  V &     \\
 &  $T_{ON}$     &     5\% - 20\% &  &     \\
 &  Lifetime     &     2 -- 5 &  years &    \\
 \hline
\end{tabular}%
}
\vspace{-3mm}
\end{table}

\noindent
\textit{(2)} \underline{{\it Testcases and architectures}}
We evaluate our carbon simulator on four industry testcases: (i) Intel server-class 2-chiplet-based  CPU, Emerald Rapids (EMR)~\cite{emr-die-area} (to be released in Q4 2023), (ii) NVIDIA GA102 GPU (2020)~\cite{ga102-die-area}, (iii) Apple A15 SoC (2021)~\cite{a15-wiki} and (iv) AR/VR (2022)~\cite{accelerator-ar-vr}. The input to our simulator is an architectural description of these testcases with the die area breakdowns for each of these processors. We obtain the area breakdowns of each of these testcases from third-party websites such as~\cite{emr-die-area, ga102-die-area, a15-die-area, tigerlake-die-area}. 

For the monolithic SoCs (GA102, and A15) we break them into chiplets based on the block-level architecture. We use one chiplet for memory, another chiplet for analog components, and a third chiplet for digital logic inspired by~\cite{AMD-Zen2}. For our 3-chiplet testcases we follow a three-tuple convention such as (7, 10, 14), which indicates the technology nodes the (digital, memory, analog components) are implemented in, respectively.   For EMR, an EMIB-based 2-chiplet testcase, we perform CFP estimation on the original architecture as is. 

% We also perform carbon evaluation on a 4-chiplet version of the same testcase where we split the large digital logic blocks into two smaller chiplets each.  

% In the rest of the results section, we use the NVIDIA GA102 testcase as a case study and show detailed results on each of the components (manufacturing, packaging, and design) of the total embodied CFP for different possible chiplet architectures and compare it to the CFP of the monolithic SoC. 
% Next, in the interest of space, for the rest of the three testcases, we only summarize the manufacturing and packaging CFP and compare that to its monolithic counterpart. 

\begin{figure*}
\centering
\includegraphics[width=0.86\linewidth]{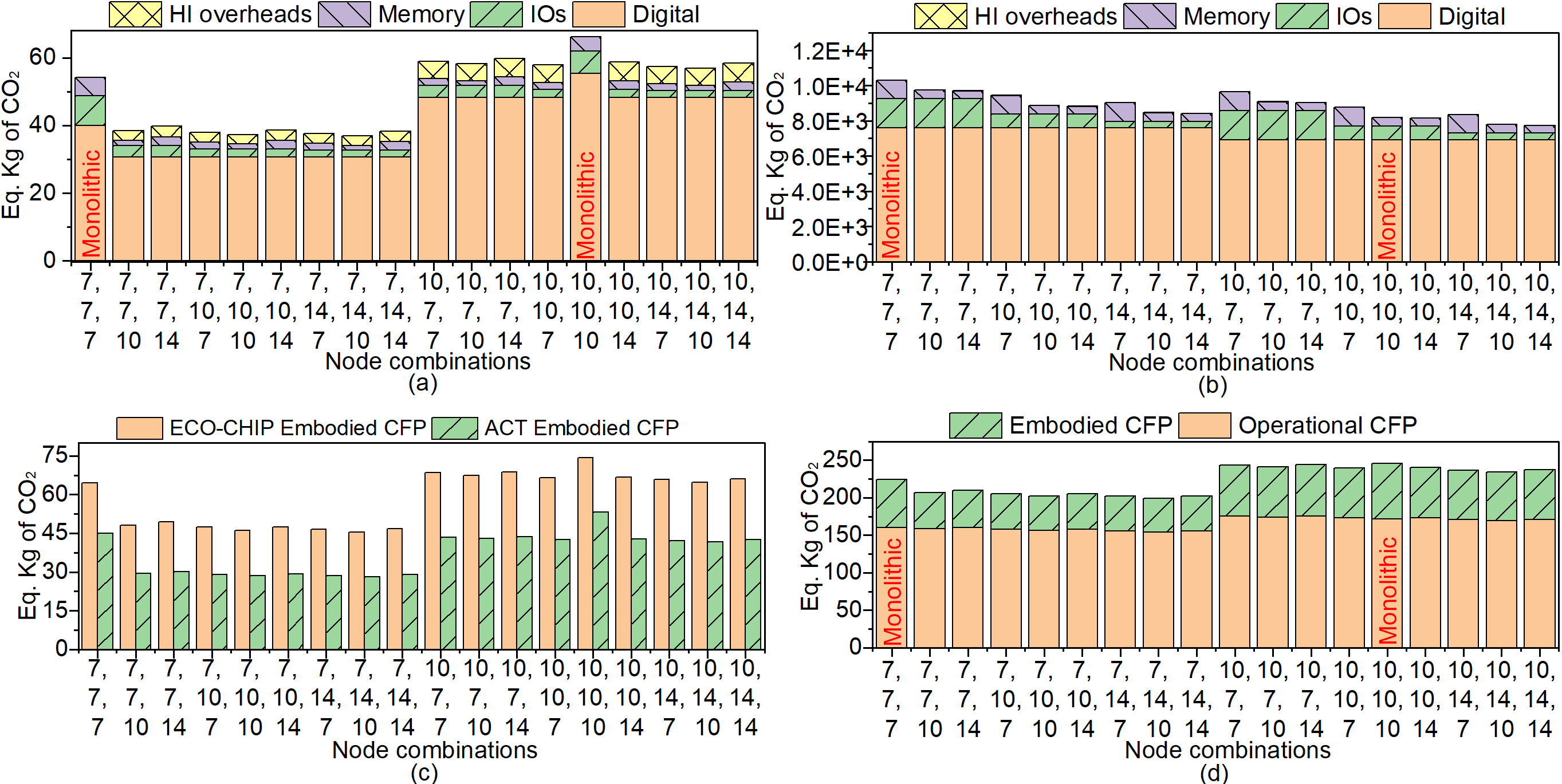}
% \vspace{-1mm}
\caption{(a) $C_{\text{mfg}}$ and $C_\text{HI}$, (b) $C_\text{des}$ for a {\bf single iteration} of SP\&R, (c) $C_\text{emb}$ for different configurations of three chiplets ($C_\text{des}$ uses $N_\text{iter} = 100$, and $N_s = 100,000$) compared with $C_\text{emb}$ from ACT~\cite{act, act+}, and (d) $C_\text{tot}$ split into its $C_\text{op}$ and $C_\text{emb}$ for the GA102 3-chiplet architecture with RDL fanout.}
\vspace{-7mm}
\label{fig:cfp}
\end{figure*}

\section{Evaluation of potential CFP savings due to HI }
\label{sec:results}
\noindent
We evaluate total CFP and highlight the new design space and CFP savings chiplet-based technologies enable through technology node mix and match, different choices of packaging architecture, and chiplet reusability. 

\subsection{Chiplet technology space exploration for reduced CFP}
\noindent
We demonstrate how the ability to mix and match technology nodes for different chiplets in a system improves embodied CFP. At the same time, we will compare ECO-CHIP embodied CFP with the existing embodied CFP estimator, ACT~\cite{act, act+}, and highlight how ACT grossly miscalculates the CFP as it does not model for package assembly, wafer area wasted, and design CFP. As an example case study, we use a 3-chiplet version of NVIDIA GPU GA102 with RDL fanout-based packaging architecture to evaluate the various components of  CFP for various chiplet disaggregation scenarios.

% \subsection{Example case study: NVIDIA GPU GA102}
% As an example case study, we use the NVIDIA GPU GA102 architecture to evaluate the various components of embodied CFP for various chiplet disaggregation scenarios of the testcase.  The first set of results is for three-chiplet scenario where one of the chiplets implements memory, the other analog components, and the third digital blocks. The next set of results will show the CFP for a general $N_c$-chiplet architecture. 

\noindent
\textit{(1)} {\underline{\it{Manufacturing and HI-related CFP}}} The manufacturing (chip and package) CFP of GA102 with RDL fanout packaging, for different configurations of technology nodes for each chiplet, is shown in Fig.~\ref{fig:cfp}(a). The x-axis lists the three-tuple configuration listing the technology node each chiplet is implemented in. The (7,7,7) scenario is a monolithic representation of the architecture of a single die in a 7nm node. It, therefore, does not have the additional HI-related packaging overheads. The figure shows that the lowest $C_\text{emb}$ is for the (7, 14, 10) scenario. This is because the analog components and memory blocks~\cite{AMD-Zen2} do not scale in the area as much as the digital blocks and can therefore be implemented in an older technology node with almost the same area. On the contrary, in the (10, 10, 10) scenario, the digital logic scales to a much larger area and therefore has a larger CFP than even the monolith resulting in a larger CFP. 

From this result, it is clear that HI enables using chiplets that have smaller areas and higher yields, which helps lowers the CFP, and the further integration of chiplets in different technology nodes can further lower the CFP as older nodes have lower $\text{EPA}$ than advanced nodes. ACT~\cite{act, act+}, uses a fixed value of package assembly CFP (150g of CO$_2$) irrespective of the area of the package, or type of packaging architecture, or the wafer area wasted and therefore can inaccurately estimate $C_\text{mfg}$ by at least 10kg of CO$_2$ emission ($\approx 20\%$ of $C_\text{emb}$).

\noindent
\textit{(2)} \underline{\it Design CFP}
From our experiments in performing SP\&R of large designs, we find that the $t_{\text{SP\&R}, i}$ for a design with 700,000 logic gates in a 7nm commercial technology is about 24 CPU hours. These estimates are on a 192GB RAM machine with a dual-core Intel Xeon CPU with 8 threads, each running at a 2.4GHz clock frequency.  Therefore, extending this model to the GA102 testcase, $t_{\text{SP\&R},i} = 1.5\times 10^5$ CPU hours as it has over 4.5B logic gates. Assuming $P_\text{des} = 10$W~\cite{power-per-core} and the energy supplied comes from non-renewable sources, then a single run of SP\&R results in 8,400kg of CO$_2$ equivalent emission in the 7nm technology node. Fig.~\ref{fig:cfp}(b), shows the design carbon for a {\bf single iteration} of SP\&R for the 3-chiplet testcase. Older technology nodes have lower design times due to EDA tool scaling~\cite{eda-tool-scaling}, and therefore, have lower CFP compared to the monolithic SoC in an advanced 7nm technology. 
In addition, since HI enables the ``reuse" of pre-designed chiplets, in principle, the same chiplet can be reused for another design saving the entire associated $C_\text{des}$.

Although the $C_\text{des}$ values in Fig.~\ref{fig:cfp}(b) are significantly large, these costs are amortized across the number of parts manufactured ($N_S$). The figure only shows the results for a single iteration of SP\&R. However, with hundreds ($N_\text{des} = 100$) of design iterations and SP\&R runs and verification dominating 80\% of the product development time, the design of an IC can easily contribute to over 2,000,000kg of CO$_2$ equivalent emission, assuming all compute energy is coming from non-renewable sources. Assuming the number of manufactured parts is $N_S=100,000$, the SP\&R carbon cost gets amortized to 12kg of CO$_2$ equivalent emission per IC, which is more than 25\% of $C_\text{mfg}$ (see Fig.~\ref{fig:cfp}(a)). This significant contributor to $C_\text{tot}$ has not been considered in ACT~\cite{act, act+}.

\noindent
\textit{(3)} \underline{\it Embodied CFP}
To estimate $C_\text{emb}$, we sum the $C_\text{mfg}$ and $C_\text{HI}$ CFP from Fig.~\ref{fig:cfp}(a) and amortized $C_\text{des}$ assuming, $N_p = 100,000$ and $N_\text{des} = 100$ from Fig.~\ref{fig:cfp}(b).  Fig.~\ref{fig:cfp}(c) shows the $C_\text{emb}$ for different configurations of the 3-chiplet GA102 testcase.  $C_\text{emb}$ is compared against ACT~\cite{act, act+} $C_\text{emb}$. Since ACT does not estimate $C_\text{des}$ and packaging-related CFP, it inaccurately estimates a lower CFP.

\noindent
\textit{(4)} \underline{\it Total CFP}
Fig.~\ref{fig:cfp}(d) shows $C_\text{tot}$ split into $C_\text{op}$ and $C_\text{emb}$ components. Given the power-hungry GPU~\cite{ga102-spec}, with a maximum power rating of 450W and average $E_{use} = 228$kWhr, the embodied carbon is approximately 20\% of $C_\text{tot}$. HI lowers the $C_\text{emb}$ compared to a monolithic SoC but increases the $C_\text{op}$ due to communication overheads and use of chiplets implemented in old technology nodes (larger supply voltages). For the GA102 testcase, the decrease in $C_\text{emb}$ dominates the increase in $C_\text{op}$ (over a  two-year lifetime), making the HI system more sustainable than the monolith.  

\begin{figure}
\centering
\includegraphics[width=0.8\linewidth]{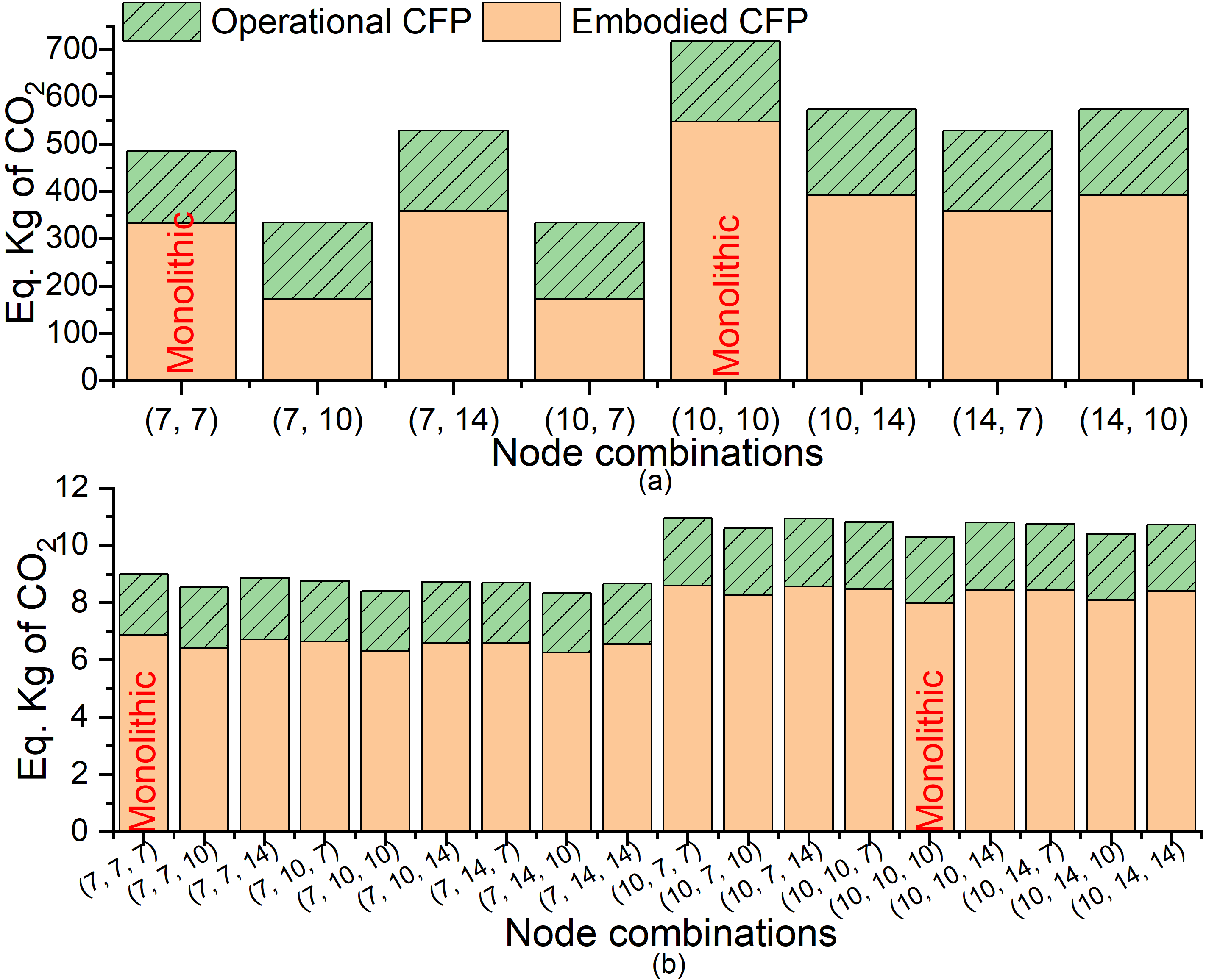}
\caption{Total CFP compared to monolithic counterparts for (a) EMR 2-chiplet with EMIB packaging (b) A15 mobile processor with RDL fanout packaging.}
\vspace{-6mm}
\label{fig:total-cpf-other-testcases}
\end{figure}

In low-power battery-operated devices, $C_\text{emb}$ dominates  $C_\text{op}$~\cite{apple-report, act}, and savings in the $C_\text{emb}$ significantly lower $C_\text{tot}$.  For example, Fig.~\ref{fig:total-cpf-other-testcases}, shows $C_\text{tot}$ split into $C_\text{op}$ and $C_\text{emb}$ for (a)~the 2-chiplet EMR testcase and (b) a 3-chiplet version of A15 mobile processor, both compared to their monolithic counterparts. For $C_\text{des}$, we assume $N_S = 100,000$ and $N_\text{iter} =~100$. The $E_\text{use}$ value is obtained from battery specification and a  battery charging rate for the mobile phone, and for the EMR testcase it is obtained by profiling a server-class CPU. The figure shows that the mobile processor has a lower operational footprint percentage (40\%), unlike the CPU/GPU processors. The improvements in $C_\text{emb}$ due to technology mix and match are lower in A15 compared to GA102, as it is smaller in area.  

\noindent
\textit{(5)} \underline{\it Key takeaways} 
(a) We find that design and package assembly CFP are significant components to total CFP and cannot be ignored as in~\cite{act,act+}. (b) Chiplets implemented in different technology nodes, lower EPA, improve yield, and provide a whole new design space to explore. (c) Similar to the insight in~\cite{pgupta-cost} concerning dollar cost, we find that larger SoCs are more suited to benefit from $C_\text{emb}$ savings when disaggregated into chiplets when compared to smaller SoCs. The $C_\text{emb}$ of GA102 lowers by 30\% when compared to its monolithic counterpart. (d) Low-power SoCs have a lower $C_\text{op}$ to $C_\text{emb}$ ratio and are more suited to benefit from a reduction in total CFP when disaggregated to chiplets.

\begin{figure}
\centering
\includegraphics[width=0.85\linewidth]{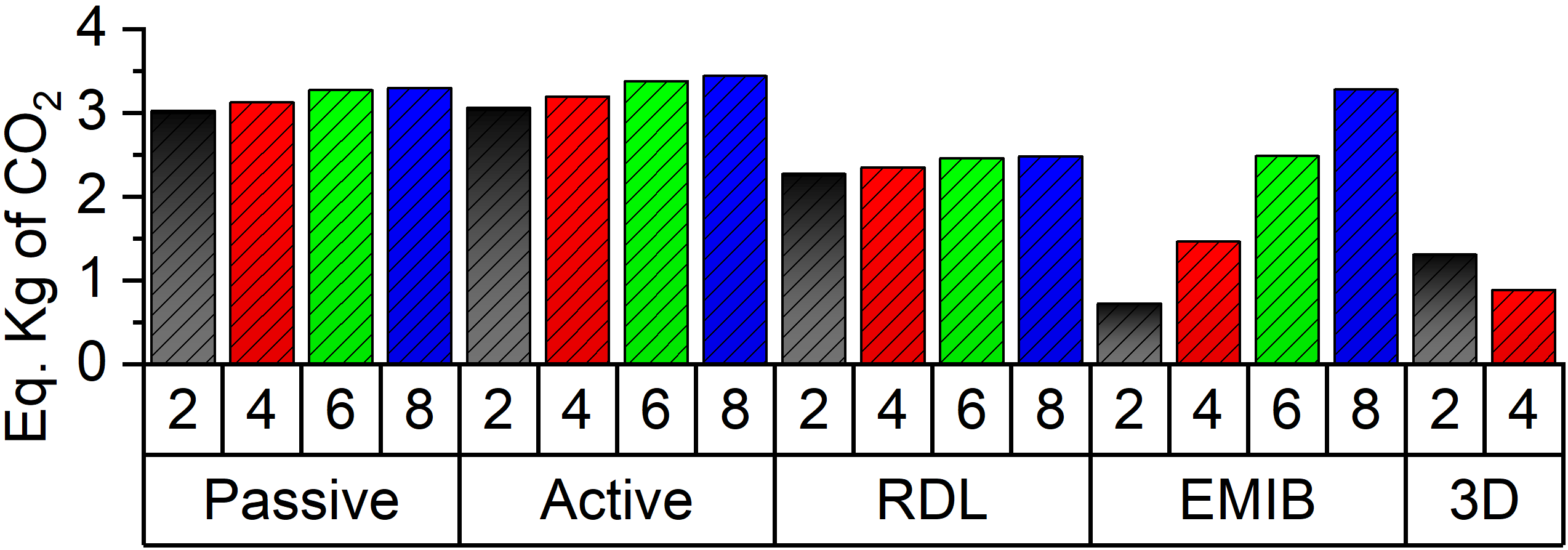}
% \vspace{-2mm}
\caption{HI-related CFP overheads for different $N_c$ values.}
\vspace{-8mm}
\label{fig:package-cfp}
\end{figure}

\subsection{Packaging technology space exploration for reduced CFP}
\noindent
\textit{(1)} \underline{\it HI-related CFP overheads for different packaging types} Although the choice of packaging architecture is driven by application requirements such as bandwidth, area, and power, the CFP for different packages varies significantly and can be considered a metric to drive early architectural decisions.  To understand the differences in CFP overheads of the five packaging architectures considered, we use the large digital logic component of GA102 as an example testcase. We split the 500mm$^2$ monolithic digital logic block into $N_c$ different chiplets and evaluate $C_\text{HI}$. Fig.~\ref{fig:package-cfp} shows the difference in CFP for these architectures separated by routing overheads and package-related overheads (whitespace and area).  All the interconnects in the package substrate are modeled in a 65nm technology for the five packaging architectures.  

Silicon-bridge-based (EMIB-based) architectures have the least CFP for 2-chiplet-based architectures of the 500mm$^2$ monolith testcase. However, as $N_c$ increases, the number of silicon bridges also increases, and CFP increases. The RDL-based packages have the least overheads for the 6- and 8-chiplet architectures, but due to their architecture definition, they have lower communication bandwidth when compared to silicon bridges or interposers. Therefore, based on the bandwidth requirements of the testcase, such tradeoffs between performance and CFP can be considered using ECO-CHIP. The figure also shows that the passive interposer has lower routing overheads as the router is part of the chiplet and is in the same technology node of the chiplet. Therefore, in passive interposer technologies, due to the advanced node (7nm in this testcase) in which the router is implemented, the area overheads are smaller than the 65nm router in the active interposer. The routing overheads of RDL, passive interposer, and silicon-bridge (EMIB) architectures are small and near-negligible compared to the core chiplet areas. For the 3D package, we sweep the number of tiers/chiplets (from two to four); as $N_C$ increases, the 2D area of each chiplet reduces, and more chiplets are stacked to implement the same logic. Therefore, the CFP decreases despite the reduction in overall package yield due to an increase in the number of TSV/bumps (the package yield is the product of the yield of each tier).

\begin{figure}
\centering
\includegraphics[width=0.75\linewidth]{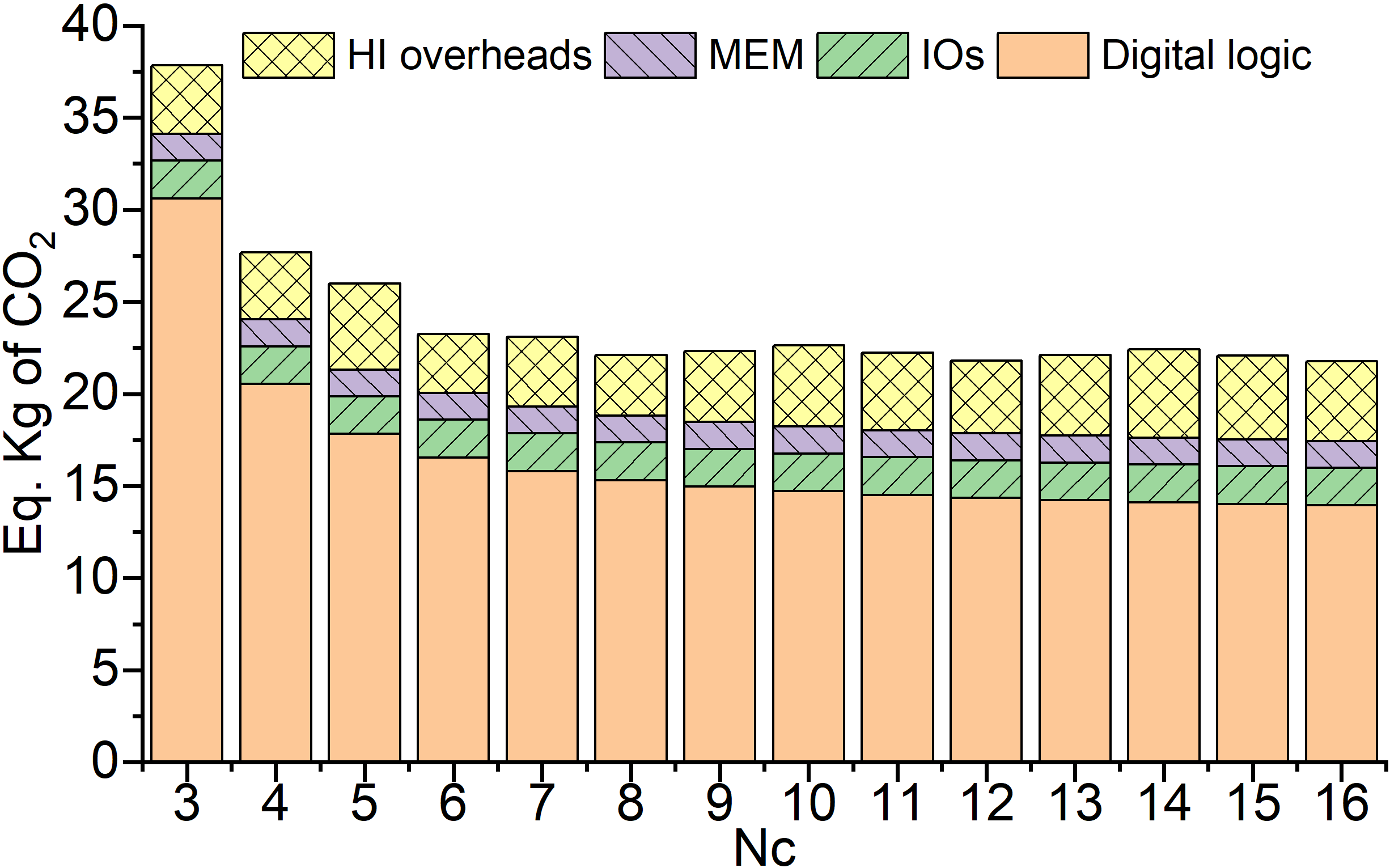}
\caption{Manufacturing CFP ($C_\text{mfg}$ and $C_\text{HI}$) for $N_c$ chiplets in GA102.}
\vspace{-8mm}
\label{fig:cfp-nc-chiplets}
\end{figure}

\noindent
\textit{(2)} \underline{\it Manufacturing CFP and HI overheads with $N_c$ chiplets} In addition to the 3-chiplet architecture of GA102, we also evaluate the $C_\text{mfg}$ and $C_\text{HI}$ for $N_c > 3$ where the digital logic block is further split it into smaller chiplets each implemented in 7nm. The analog (IOs) and memory chiplets are in 14nm and 10nm, respectively. Fig.~\ref{fig:cfp-nc-chiplets} shows $C_\text{mfg}$ and $C_\text{HI}$ for different $N_c$. As $N_c$ increases, the $C_{\text{mfg},i}$ decreases due to smaller chiplets and better yields, while $C_\text{HI}$  overheads increase. The data shows that beyond a certain chiplet size (or $N_C$), the CFP savings reduces as $C_\text{HI}$ dominates.

\begin{figure}
\centering
\includegraphics[width=0.9\linewidth]{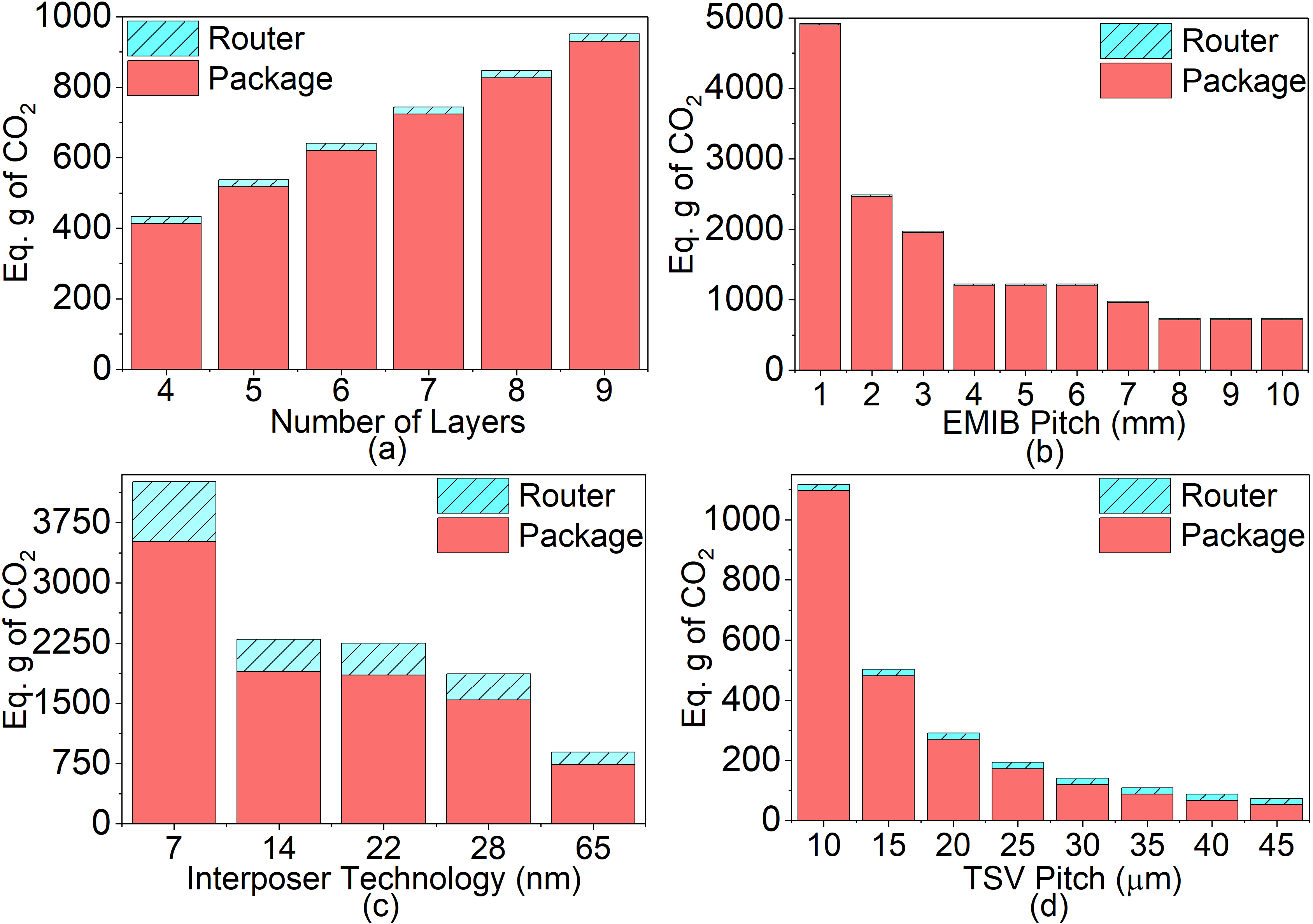}
%\vspace{-3mm}
%\vspace{-1mm}
\caption{ $C_\text{HI}$ for the A15 testcase with different parameter sweeps. $C_\text{HI}$ for different: (a) $L_\text{RDL}$ for RDL-fanout, (b)bridge ranges for EMIB, (c) interposer technology nodes for active interposer, and (d) TSV pitches for 3D.}
%\vspace{-2mm}
\label{fig:package-parameter-sweep}
\end{figure}

\noindent
\textit{(3)} \underline{\it Packaging technology parameters and its impact on CFP} 
$C_\text{HI}$ for the different packaging architectures supported by ECO-CHIP are based on the estimated area overhead and the computed package yield. For each packaging architecture, certain key parameters determine their assembly CFP. For instance, $L_\text{RDL}$ directly affects $C_\text{RDL}$ (Eq.~\eqref{eq:rdl-cfp})  as shown in Fig.~\ref{fig:package-parameter-sweep}(a). The figure sweeps the number of BEOL layers in the RDL fanout package from 4 to 9, showing the linear increase in $C_{HI}$. Fig.~\ref{fig:package-parameter-sweep}(b) shows the decrease in $C_\text{HI}$ with an increase in EMIB range/pitch. The increase in range reduces the number of bridges needed for inter-die communication lowering $C_{HI}$. Fig.~\ref{fig:package-parameter-sweep}(c) shows the difference in HI-related overheads for active interposers implemented in different technology nodes. Older technology nodes have lower EPA and therefore, lower CFP.   Fig.~\ref{fig:package-parameter-sweep}(d) sweeps the TSV pitch. Larger TSV pitches imply fewer TSVs between the two tiers ($N_{TSV, bump}$), and larger yields lowering CFP compared to smaller TSV pitches.

\begin{figure}
\vspace{-3mm}
\centering
\includegraphics[width=\linewidth]{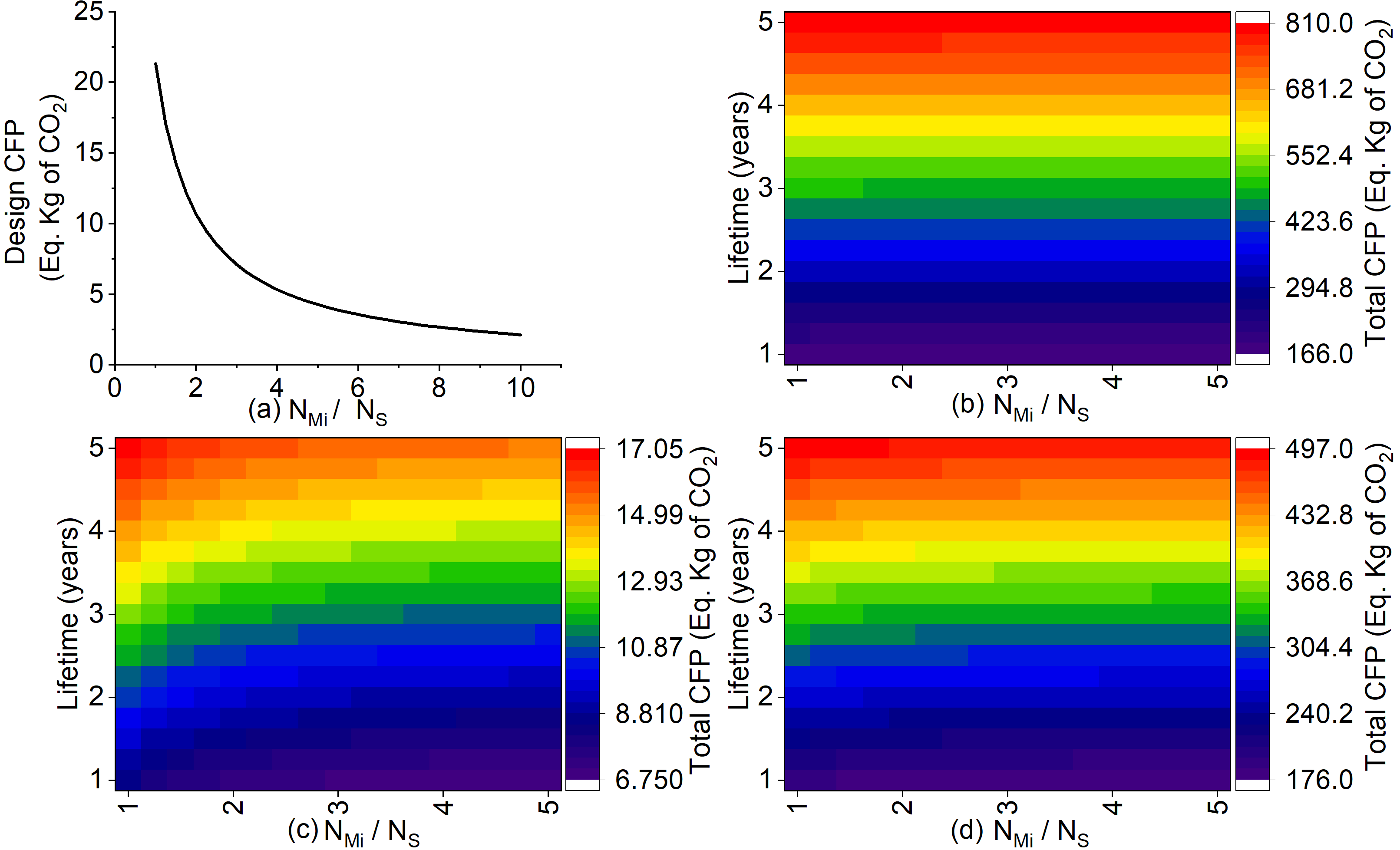}
\vspace{-5mm}
\caption{ (a) $C_\text{des}$ vs. $\frac{N_{M_i}}{N_S}$ highlighting the reduction in $C_\text{des}$ with the increase in manufacturing volume. Variation in $C_\text{tot}$ as a function of $\frac{N_{M_i}}{N_S}$ and lifetime for (b) GA102, (c) A15, and (d) EMR 2-chiplet testcases in 7nm. }
\vspace{-6mm}
\label{fig:heat-maps}
\end{figure}

% \begin{figure*}[t]
%   % \vspace{-0.2em}
%   \centering
%   \begin{minipage}{.3\textwidth}
%   \includegraphics[width=\linewidth]{figs/carbon-dpa-curve.png}
%    \vspace{-4mm}
%   \caption{ \redHL{(a) Carbon-delay product curve, (b) Carbon-power product curve, and (c) Carbon-area product curve for the accelerator testcase~\cite{accelerator-ar-vr}.}}
%   \vspace{-7mm}
%    \label{fig:accelerator-cfp}
%   \end{minipage} \hfill
%   \begin{minipage}{.3\textwidth}
% %\includegraphics[width=\linewidth]{figs/ga102-cost-impt-comb.png}
% \includegraphics[width=\linewidth]{figs/ga102-cost-2.png}
% \vspace{-3mm}
% \caption{Dollar cost variation for (a) different technology node combinations for 3-chiplet GA102 testcase, (b) disaggregating the GA102 testcase into $N_C$ chiplets~\redHL{(1) Chetan, can you reduce the height of these plots? The columns are unnecessarily tall, you can have lesser space between the ticks on the y axis. Both these figures should have the same height of the figure on the left. (2) I agree about not much variation in cost in fig (b). But this might be good. Because, for the same cost we can get a significantly lower CFP from Fig. 7. So might be reasonable to show}}
% \vspace{-2mm}
% \label{fig:ga102-cost}
%   \end{minipage} 
% % \vspace{-1.em}
% \end{figure*}

\subsection{Chiplet reusability for reduced CFP}
\noindent
Besides the ability to ``mix and match" technology nodes for different chiplets and the improvements in yield with a HI system, the ability to reuse chiplets also helps lower $C_\text{emb}$ as the $C_\text{des}$ and the NRE component of the $C_\text{mfg}$ is amortized across the number of chiplets manufactured ($N_{M_i}$) and used in a variety of systems in different applications. Several standard IP blocks such as USB, PCIe etc.,  can be manufactured in large volumes as chiplets that can then be used across several different systems amortizing $C_\text{des}$. Further, when chiplets are manufactured in large volumes, the CFP associated with NRE costs, such as manufacturing and designing the masks used during photolithography, also gets amortized across $N_{M_i}$. Although ECO-CHIP does not split the $C_\text{mfg}$ into its NRE and non-NRE components, this will only improve CFP savings. 

Fig.~\ref{fig:heat-maps}(a) shows the sweep of the ratio of $N_{M_i}$ to $N_{S}$ for the EMR 2-chiplet testcase and plots $C_\text{des}$ (with $N_\text{iter} = 100$, $N_{M_i} = 100,000$)(Refer Eq.~\eqref{eq:des-cfp}) when both chiplets are implemented in 7nm technology node. Larger $\frac{N_{M_i}}{N_S}$ results in lower $C_\text{des}$ as the cost of designing the chiplet is amortized across a larger number of systems.  The figure shows the potential $C_\text{des}$ savings from the reusability of chiplets, i.e., chiplets can be designed once and reused several times.

Fig.~\ref{fig:heat-maps}(b), (c), and (d) shows $C_\text{tot}$ across a lifetime of 5 years for different $\frac{N_{M_i}}{N_S}$ ratios for the GA102 and A15 with RDL fanout-based packaging, and EMR with EMIB-based packaging testcases, respectively. With the increase in the ratio, the $C_\text{emb}$ reduces, and with an increase in lifetime, the $C_\text{op}$ increases. On one hand, in GA102 (Fig.~\ref{fig:heat-maps}(b)), the $C_\text{op}$ dominates $C_\text{emb}$ and therefore, $C_\text{tot}$ does not reduce significantly with the increase in the ratio. On the other hand, in the A15 testcase (Fig.~\ref{fig:heat-maps}(c)), where the $C_\text{emb}$ dominates the $C_\text{op}$, increasing the ratio helps lower $C_\text{tot}$. Therefore, in A15, reducing $C_\text{tot}$ requires reducing the $C_\text{emb}$, especially given the lifetime of these consumer mobile processors is small. 
With a $N_{M_i}$ value of 100,000, used across all testcases, the figure indicates how many systems each chiplet must be utilized in, to amortize the embodied cost across the lifetime of operation. This analysis is helpful when determining the required volumes in which chiplets much be manufactured.

\begin{figure}
\centering
\includegraphics[width=\linewidth]{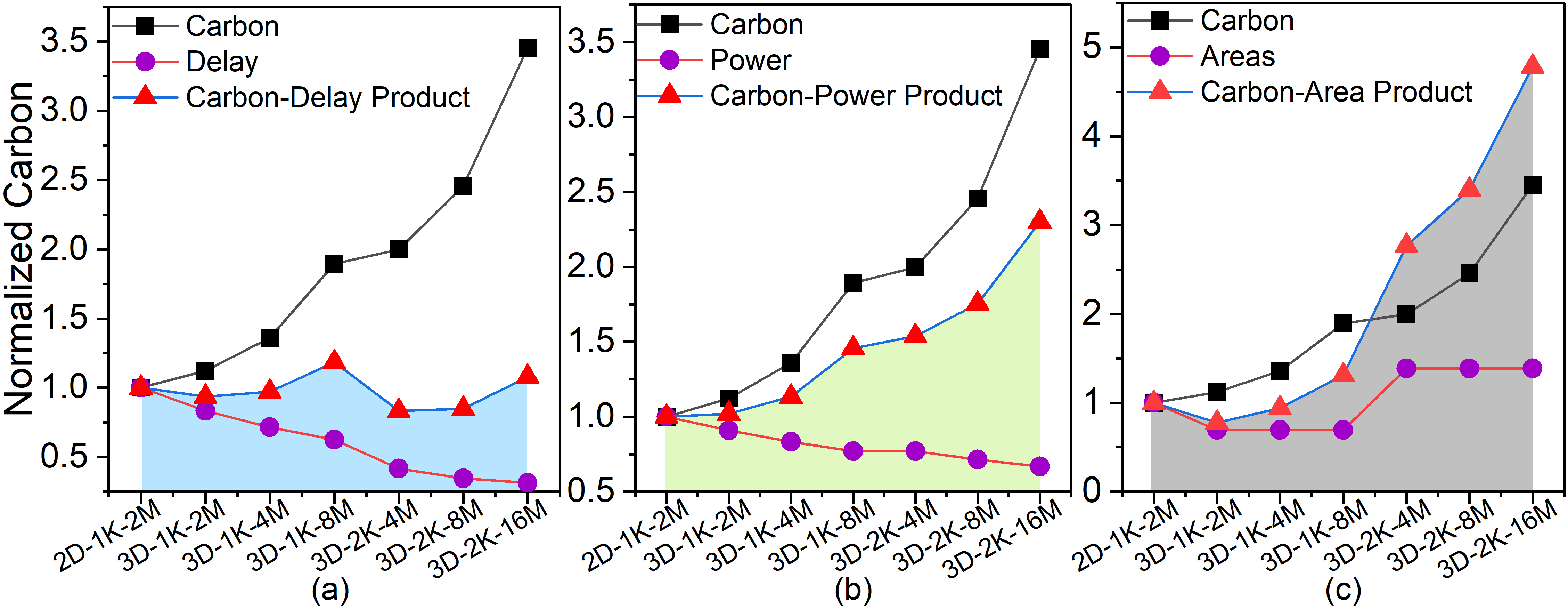}
\vspace{-3mm}
\caption{ (a) Carbon-delay product curve, (b) carbon-power product curve, and (c) carbon-area product for different 3D configurations of the accelerator.}
\vspace{-6mm}
\label{fig:accelerator-cfp}
\end{figure}

\vspace{-1mm}

\section{Design and architecture space exploration for chiplet disaggregation}
\label{sec:dse}
\noindent
In this section, we demonstrate the application of ECO-CHIP in performing design space exploration, considering CFP as a first-order optimization metric along with performance, power, area, and cost. To analyze the CFP tradeoffs with these metrics, we consider the accelerator testcase described for AR/VR applications~\cite{accelerator-ar-vr} in addition to three testcases in Section~\ref{sec:results}. The testcase uses a 3D packaging integration technique with 1-4 SRAM dies stacked on top of a computation unit using microbumps in a 7nm technology. The testcase comes in two flavors. The first, 1K, uses SRAM dies of 2MB capacity each, and the second, 2K, uses SRAM dies of 4MB capacity each. The naming convention for each of these testcases is as follows: 2D/3D-1K/2K-2MB/4MB/8MB/16MB. For instance, a 3D-1K-4M is a 3D architecture with 2 tiers of a 2MB SRAM chiplet and a total memory of 4MB. 

\noindent
\textit{(1)} \underline{\it Delay, power, area, and CFP tradeoffs}
Fig.~\ref{fig:accelerator-cfp}(a) shows the carbon-delay product curve for the accelerator testcase with different numbers of SRAM tiers. As SRAM tiers increase (from left to right in the figure) for each 1K or 2K series, the system latency reduces, but ECO-CHIP finds that $C_\text{tot}$ increases.  $C_\text{tot}$ is estimated for a lifetime of 2 years, where $C_\text{op}$ is estimated using $E_\text{use}$ provided in~\cite{accelerator-ar-vr}.  ECO-CHIP shows that as the number of tiers increases, although the delay improves, the embodied $C_\text{emb}$ increases as there is an increase in total memory capacity and silicon dies. 

Fig.~\ref{fig:accelerator-cfp}(b) and (c) show the carbon-power and the carbon-area product curves for the same accelerator testcase. The curves show that as the number of SRAM tiers increases, the energy efficiency of the accelerator improves, reducing operational power~\cite{accelerator-ar-vr} and carbon. However, since the $C_\text{emb}$ dominates, the $C_\text{tot}$ increases as the number of tiers increases.  Since it's a 3D system, the 2-dimensional area of each configuration remains within 1K or 2K.  These product curves enable design space exploration, allowing the selection of an architecture that meets the latency, power, and area specifications while minimizing $C_\text{tot}$.

Since the performance of the HI system is very application- and testcase-specific, estimating the performance overheads of the chiplet-based GA102 testcase and A15 testcase, which are originally monolithic, requires modeling the performance of inter-die communication and router overheads, which is beyond the scope of ECO-CHIP. Therefore, for performance and CFP tradeoff curve, we only consider the accelerator testcase with delay numbers that are readily available~\cite{accelerator-ar-vr}. 

\begin{figure}
\centering
\includegraphics[width=0.75\linewidth]{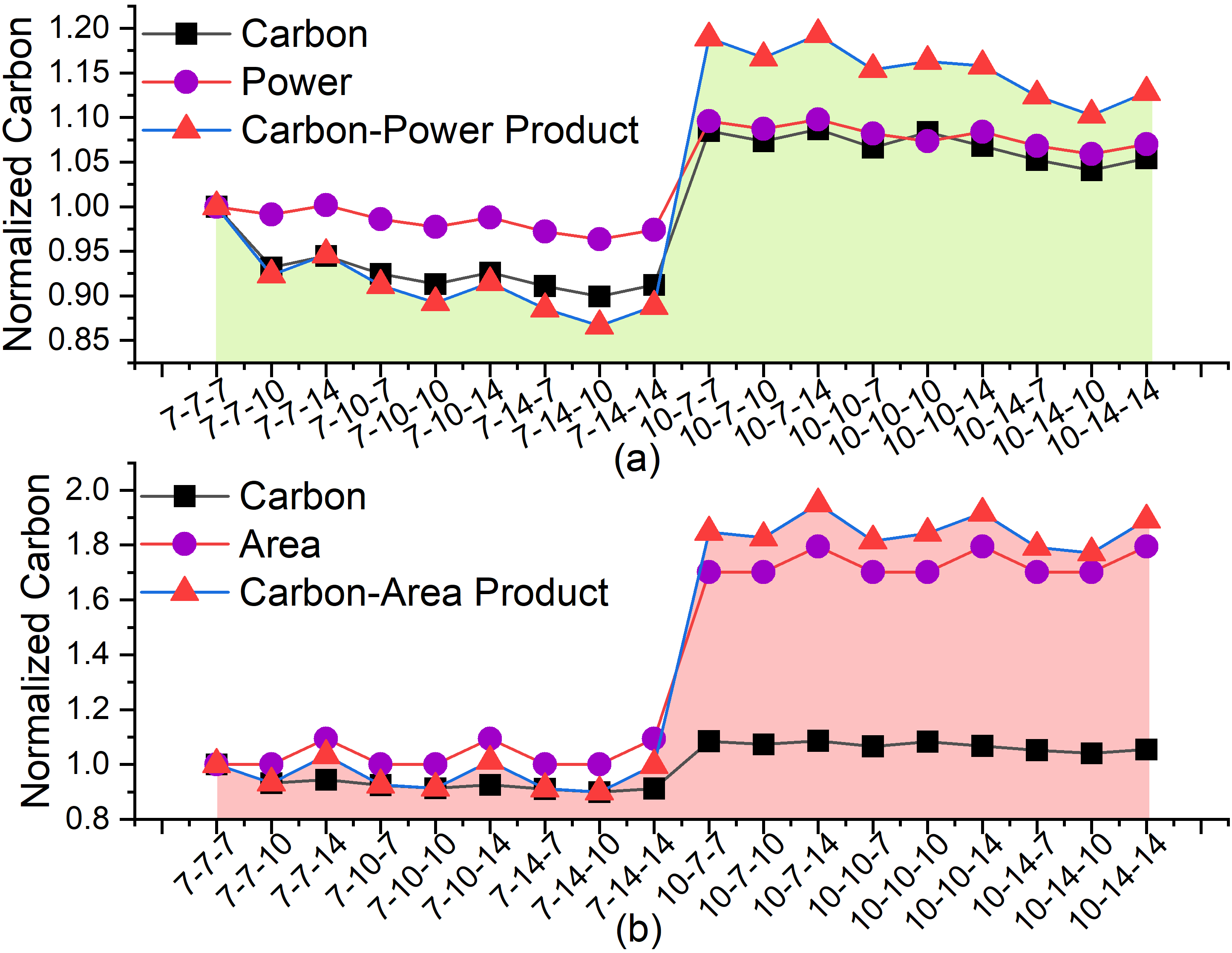}
%\vspace{-1mm}
\caption{(a) Carbon-power product and (b) carbon-area product for GA102.}
\vspace{-4mm}
\label{fig:ga102-cfp-power-area}
\end{figure}

Fig.~\ref{fig:ga102-cfp-power-area}(a) and (b) show the variation in operational power and area and total CFP product for the GA102 3-chiplet with RDL fanout package testcase for different technology nodes normalized to its monolithic counterpart. Older technology node chiplets have larger chip areas and power due to HI-related overheads such as white space on the substrate/interposer, additional router logic etc. However, older technology nodes have a lower CFPA lowering $C_\text{tot}$ (Fig.~\ref{fig:cfp}). ECO-CHIP enables considering these tradeoffs, to drive decisions related to SoC disaggregation into chiplets.

\begin{figure}
\centering
\includegraphics[width=0.9\linewidth]{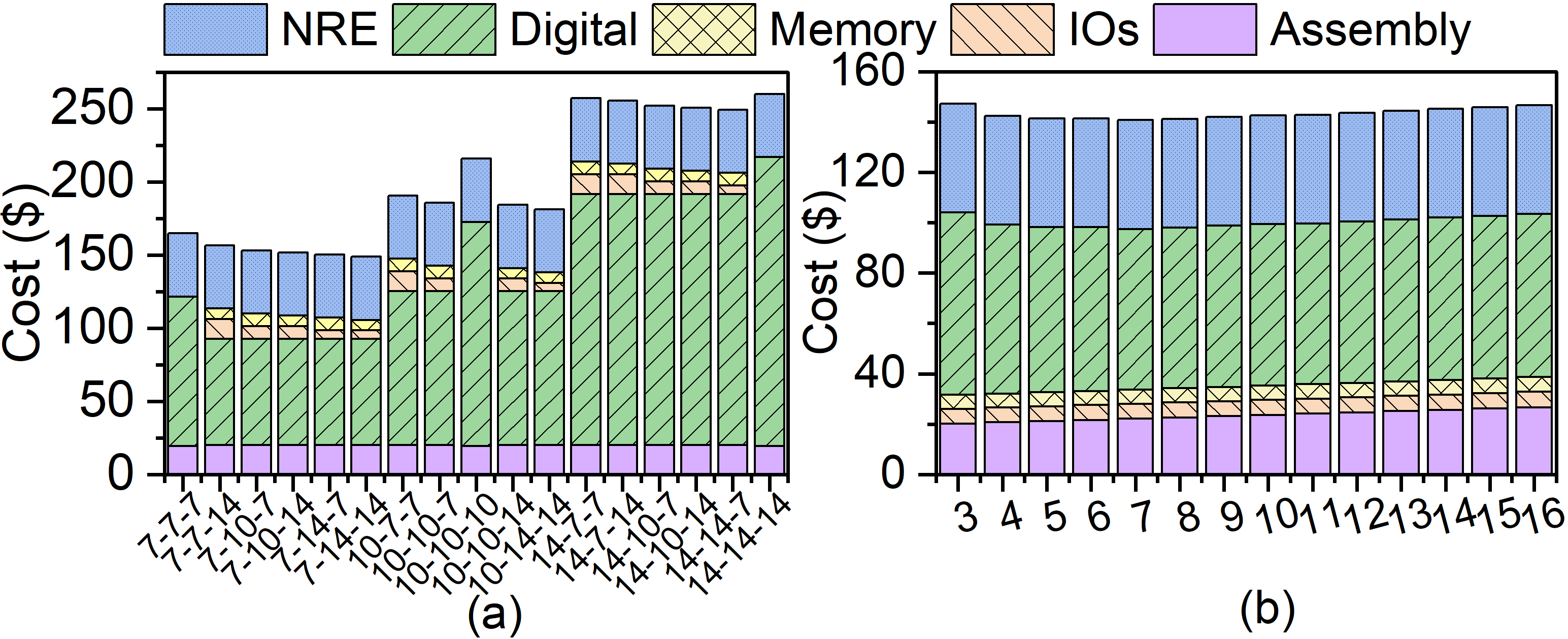}
\vspace{-1mm}
\caption{ Cost variation for (a) different technology node configurations of GA102 testcase, (b) disaggregating the GA102 testcase into $N_C$ chiplets.}
\vspace{-6mm}
\label{fig:ga102-cost}
\end{figure}

\noindent
\textit{(2} \underline{\it Dollar cost analysis}
ECO-CHIP integrates with a third-party chiplet-based dollar cost analysis tool~\cite{pgupta-cost} and uses default parameters in~\cite{pgupta-github} for cost estimation. We input the architectural description (areas and technology nodes) of our testcase with identical yield numbers used for CFP estimation. Fig.~\ref{fig:ga102-cost}(a) shows the dollar cost associated with the 3-chiplet GA102 testcase for different technology node combinations. The dollar cost follows a similar trend as the total CFP trend in Fig.~\ref{fig:cfp}(d), where older technology node chiplets have lower costs due to better yields and cheaper manufacturing.

Fig.~\ref{fig:ga102-cost}(b) shows the variation in dollar cost as we split the GA102 digital logic block into $N_C$ chiplets similar to manufacturing CFP cost in Fig.~\ref{fig:cfp-nc-chiplets}. The assembly cost increases as the number of chiplets increases and the cost of manufacturing the digital logic block decreases due to an increase in yield of smaller die sizes as the number of chiplets increases. The cost variation in Fig.~\ref{fig:ga102-cost}(b) is small compared to the variation in $C_\text{HI}$ and $C_\text{mfg}$ in Fig.~\ref{fig:cfp-nc-chiplets}, which allows an architect to consider the $N_C$ with the least $C_\text{tot}$ from a cost perspective. 

\section{Validation Discussion}
\noindent
It's important to note here that ECO-CHIP is a tool to perform analysis on embodied and operational CFP of heterogenous (HI) systems which has not been done before. It's a methodology that is available open-source and can generate numbers as accurate as the accuracy of the input parameters, e.g., design time, yields, and defect densities, and is easily adoptable by the industry that has access to accurate numbers. ECO-CHIP is based on CFP data from two sources. First, ~\cite{imec-dtco} which provides manufacturing CFP numbers from IMEC reported on a per metal layers basis for different technology nodes. Second, ACT~\cite{act}, which provides manufacturing CFP numbers mined from industrial sustainability reports ~\cite{apple-report, TSMC-report, microsoft-report, facebook-report, spil-packaging, amkor-report, ace-report}, and a carbon footprint per unit area. The HI-related CFP estimation also relies on the same CFP numbers and is estimated by modeling additional area overheads and metal layers in interposer/package substrate for each kind of packaging architecture.  As a sanity check of the CFP numbers generated by ECO-CHIP, we compare our A15 processor CFP numbers with that reported in Apple’s sustainability report for the entire iPhone 14. We find that our reported numbers are approximately 16\% of the total CFP of the iPhone. As reported by the Apple sustainability and product report ~\cite{apple-report2023,iphone14-report}, ECO-CHIP also estimates 20\% of the total CFP is for operational CFP and 80\% for embodied CFP as seen in Fig.~\ref{fig:total-cpf-other-testcases}(b). Validation is indeed a challenging problem, especially given the coarse granularity at which industry sustainability reports are provided and the lack of open-source data from the industry on various input parameters such as design time, yields, etc. For example, Apple provides CFP for the iPhone as a whole and it's difficult to figure out the contributions of the A15 processor alone. However, the industry that has accurate data on yields, design time, etc., can utilize ECO-CHIP to generate accurate CFP. 

\section{Related work}
\label{sec:related-work}
\noindent
Two prior bodies of work focus on CFP estimation at the architectural level: the first body of work includes~\cite{firstorder-simple, kaya-ghent}, and the second includes~\cite {chasing-carbon, act, act+}.  The work in~\cite{kaya-ghent} reformulated the Kaya identity~\cite{kaya-env} to understand how the global CFP of computer systems evolves and has made a case to lower chip sizes to lower embodied CFP and~\cite{firstorder-simple} creates a simple model based on first principles. The works in~\cite{act, chasing-carbon, act+} have created data-driven model, from publicly available sustainability reports from industry~\cite{apple-report, TSMC-report, imec-wp, facebook-report}, for embodied carbon estimation and have created a platform for carbon-aware design space exploration (DSE)~\cite{act+}. While these works have set a new paradigm, they are limited in scope:
\begin{enumerate}
     \item They do not apply to emerging HI systems where small chiplets are integrated into a single package.
    \item They do not accurately consider the packaging/assembly carbon costs, which is crucial for HI systems. ACT~\cite{act} uses a fixed package CFP value irrespective of the size of the package, the yield of the package, and the assembly process and~\cite{kaya-ghent} does not consider or separate the packaging CFP component.
    \item They do not consider the CFP from the {\it design} of chips, which, even though amortized across all manufacturing parts, significantly contributes to the embodied CFP. 
    \item They do not take into account silicon wastage from the periphery of the wafer (Fig.~\ref{fig:area-wasted}).
\end{enumerate} 

In contrast, and complementary, to the above two bodies of work ECO-CHIP focuses on evaluating the potential of chiplet-based systems towards sustainable computing by modeling CFP from advanced packaging architectures, yields, area scaling models, and design CFP. Our comparisons to~\cite{act, act+} in Section~\ref{sec:results} have shown the importance of considering these models in the context of HI. Unlike~\cite{act, act+}, ECO-CHIP is integrated with emerging design methodologies (such as system disaggregation) for chiplet-based systems to make them cognizant of sustainability (Refer Section~\ref{sec:dse}). 

% While analyzing the dollar cost for advanced packaging technologies is common~\cite{chopin, pgupta-cost}, there is limited work for CFP. Prior package carbon analysis tools have been limited to only traditional ball grid arrays and monolithic flip-chip technologies~\cite{lca-package} and have not been studied in the context of sustainable computing or HI. 

\section{Conclusion}
\label{sec:conclusion}
\noindent
In this paper, we proposed HI as a path towards sustainable computing by designing and manufacturing chiplet-based systems with lower embodied carbon footprint (CFP) than monolithic SoCs. We developed ECO-CHIP, a CFP estimator that uses architectural-level descriptions to assess heterogeneous systems' total CFP (embodied and operational), including advanced packaging CFP overheads. We demonstrated the use of ECO-CHIP to guide system disaggregation and design space exploration in Section~\ref{sec:dse} and integrated with other chiplet-based cost estimation tools. ECO-CHIP is open-source and available at anonymous repository~\cite{github}. We believe that ECO-CHIP will enable the development of more sustainable design methodologies for emerging heterogeneous systems.

%%%%%%%%% -- BIB STYLE AND FILE -- %%%%%%%%
\newpage
\bibliographystyle{IEEEtran}
\bibliography{bibfile}
%%%%%%%%%%%%%%%%%%%%%%%%%%%%%%%%%%%%
\newpage
\appendix
\section{Artifact Appendix}
%%%%%%%%%%%%%%%%%%%%%%%%%%%%%%%%%%%%%%%%%%%%%%%%%%%%%%%%%%%%%%%%%%%%%
\subsection{Abstract}
\noindent
The paper introduces ECO-CHIP, a framework for measuring the carbon footprint (CFP) of a heterogeneous system across its lifespan. 
This artifact is released on Zenodo and contains two parts. The first is ECO-CHIP submodule from GitHub, and the second is a folder that consists of the experiments performed using ECO-CHIP. This appendix describes the installation of our artifact, ECO-CHIP, and the procedure to reproduce the results in the paper. The minimal hardware requirements are any single-core CPU, and the software requirements are Python 3.8, python3.8-venv, with pip 20.0.2. 

\vspace{-1mm}

\subsection{Artifact check-list (meta-information)}

{\small
\begin{itemize}
  \item {\bf Algorithm: } Heterogeneous chiplet based carbon analysis tool that can evaluate the sustainability potential of Heterogeneous systems, considering scaling, chiplet, packaging yields, design complexity, and advanced packaging overheads.
  \item {\bf Program: } ECO-CHIP and the experiments are setup in Python.
  \item {\bf Compilation: } Python compiler
  \item {\bf Data set: } There is no large particular dataset. However, in our ECO-CHIP GitHub repository~\cite{github} and the Zenodo release~\cite{zenodo}, we have configuration files and architectural descriptions of the testcases used in the paper, which serve as input to ECO-CHIP in JSON format.  
  \item {\bf Run-time environment: } Runtime is not critical for ECO-CHIP. The script performs very simple equation-based calculations.  \begin{itemize}
      \item Not OS-Specific
      \item Dependencies - Python 3.8, python3.8-venv, pip 20.0.2
      \item No need for root access
  \end{itemize}
  \item {\bf Hardware: } No specific hardware requirement, at least one CPU core. 
  \item {\bf Run-time state: } Not sensitive to run-time state.
  \item {\bf Execution: } Full execution should take 10sec based on input parameters. 
  \item {\bf Metrics: } ECO-CHIP simulator estimates equivalent CO$_\text{2}$ emissions. The output will be the total CFP for the input testcase, including embodied CFP (design and manufacturing) and operational CFP. 
  \item {\bf Output: } 
  \begin{itemize}
      \item ECO-CHIP will output the CO$_\text{2}$ emission values across different combinations of technology nodes. 
      \item Will provide a breakdown of CO$_\text{2}$ values across different chiplets.
      \item Provides design and manufacture (embodied), operational, and total CO$_\text{2}$ emission values.
  \end{itemize}
  \item {\bf Experiments: } The experimental setup that we release on Zenodo (artifact) generates the key results of the paper (Fig.~\ref{fig:cfp},~\ref{fig:total-cpf-other-testcases},~\ref{fig:package-cfp},~\ref{fig:cfp-nc-chiplets},~\ref{fig:accelerator-cfp}, etc.). The experimental setup is specific to the results in the paper, where the testcase sweeps of various input parameters, etc., are present in the specific directory related to the figure to help with easy reproducibility. ECO-CHIP GitHub is set up to run different new testcases (not just the ones in the paper) and estimate CFP of the system. We provide the experiment scripts and detailed descriptions for running them. All instructions are provided in \texttt{README.md} file under \texttt{artifact} available on Zenodo (Please note that this is not available on GitHub). 
  \item {\bf How much disk space required?: } Less than 1GB
  \item {\bf How much time is needed to prepare workflow (approximately)?: } Less than 1 minute
  \item {\bf How much time is needed to complete experiments (approximately)?: } Less than 10 minutes 
  \item {\bf Publicly available?: } Yes. 
  \begin{itemize}
      \item \url{https://github.com/ASU-VDA-Lab/ECO-CHIP}
      \item \url{https://zenodo.org/records/10223759} 
  \end{itemize}  
  \item {\bf Code licenses (if publicly available)?: } BSD 3-Clause ``New" or ``Revised" License
  % \item {\bf Workflow framework used?: } Python
  \item {\bf Archived (provide DOI)?: } \begin{itemize}
      \item \url{https://github.com/ASU-VDA-Lab/ECO-CHIP}
      \item Zenodo DOI : 10.5281/zenodo.10099731
  \end{itemize} 
\end{itemize}
}
\vspace{-2mm}

%%%%%%%%%%%%%%%%%%%%%%%%%%%%%%%%%%%%%%%%%%%%%%%%%%%%%%%%%%%%%%%%%%%%%
\subsection{Description}
\noindent
\subsubsection{How to access}
The artifact to regenerate all the results in the paper is available on open-source Zenodo~\cite{zenodo}. ECO-CHIP simulator is available on GitHub~\cite{github}. 

\noindent
\subsubsection{Hardware dependencies} A CPU with at least one core.

\subsubsection{Software dependencies}
The artifact and the tool are implemented in Python 3.8 and require several packages that help run the tool. A full detailed list of required packages is in \texttt{requirments.txt} file. The \texttt{requirments.txt} is available in the repository. 

\subsubsection{Data sets}
Our GitHub repository~\cite{github} contains testcases that were used in paper, most of the files are in JSON format. Detailed descriptions of each of these files are provided in \texttt{README.md} file under the repository. The \texttt{architecture.json} contains high-level architecture details of each chiplet and the packaging type, \texttt{designC.json} contains input parameters needed for design CFP, \texttt{node\_list.txt} specifies the technology nodes of interest for CFP exploration, \texttt{operationalC.json} specifies details about the lifetime and \texttt{packageC.json} has specific parameters related to packaging. The description of each testcase in our dataset is provided in~\cite{github}.

%%%%%%%%%%%%%%%%%%%%%%%%%%%%%%%%%%%%%%%%%%%%%%%%%%%%%%%%%%%%%%%%%%%%%
\subsection{Installation}
\noindent
The installation for ECO-CHIP simulator and all the experiments in the artifact repository are the same. Once the zip file is downloaded from Zenodo~\cite{zenodo} or cloned from GitHub~\cite{github}, it requires creating a virtual Python environment to install all the packages via pip. Based on the following instructions:
\begin{itemize}
    %\item \texttt{git clone https://githusb.com/ASU-VDA-Lab/ECO-CHIP.git}
    \item \texttt{cd ECO-CHIP-AE} or \texttt{cd ECO-CHIP}
    \item \texttt{python3 -m venv eco-chip}
    \item \texttt{source eco-chip/bin/activate}
    \item \texttt{pip3 install -r requirements.txt}
\end{itemize}
The \texttt{source eco-chip/bin/activate} assumes using bash shell on the Unix environment. 

%%%%%%%%%%%%%%%%%%%%%%%%%%%%%%%%%%%%%%%%%%%%%%%%%%%%%%%%%%%%%%%%%%%%%
\subsection{Experiment workflow}
\noindent
{\bf Experiments to regenerate results in the paper}
The artifact directory available on Zenodo contains scripts and details on regenerating the results in the paper. The README file in each folder details how to run each experiment. To run all experiments, after installation (as described in the top-level README), from the artifact folder, run the bash script \texttt{run\_all.sh} in the virtual environment. A specific experiment can be run from its unique folder.  For example, 
    \begin{itemize}
        \item \texttt{cd artifact/fig2}
        \item \texttt{python3 fig2a.py}
    \end{itemize}
The result with the plot generated will be created in the \texttt{artifact/result\_img/} directory.

\noindent
{\bf ECO-CHIP simulator stand-alone}
With the simulator~\cite{github}, we provide multiple examples of testcases to measure the CFP of different heterogeneous systems and perform CFP estimation across different technology node combinations. For example, after installation, from the ECO-CHIP folder, the following command will run the GA102 testcase:
\begin{itemize}
    \item \texttt{python3 src/ECO\_chip.py --design\_dir testcases/GA102/}.
\end{itemize}
  We have multiple other testcases under the \texttt{ECO-CHIP/testcases} directory.
 
\noindent
{\bf Running ECO-CHIP for a new design}
To run ECO-CHIP on a new design, create a new directory under \texttt{ECO-CHIP/testcase/} and add the parameters of the design of interest into the JSON files. Detailed description for each of the input parameter files is provided in \texttt{ECO-CHIP/README.md} and then run the above command with a pointer to the specific design directory.

\vspace{-1mm}
%%%%%%%%%%%%%%%%%%%%%%%%%%%%%%%%%%%%%%%%%%%%%%%%%%%%%%%%%%%%%%%%%%%%%
\subsection{Evaluation and expected results}

\noindent
All the required scripts to help reproduce the key results and contributions are under the \texttt{artifact} directory. Detailed steps on how to run each of the scripts and Python packages that need to be installed have been mentioned in the \texttt{README.md} file. The results of our artifact will be reproducible and may have very small variations in single-digit percentages. All the graphs generated by the script will be under \texttt{artifact/result\_img/} directory. Running the scripts such as \texttt{fig7.py, fig8a.py, fig8b.py, fig9.py, fig10.py, and fig13.py} will help generate the respective plots under the \texttt{artifact/result\_img/} directory. This can verify the critical result of the paper as shown in Fig.~\ref{fig:cfp}, Fig.~\ref{fig:total-cpf-other-testcases}, Fig.~\ref{fig:package-cfp}, Fig.~\ref{fig:cfp-nc-chiplets}, and Fig.~\ref{fig:accelerator-cfp}, as the main contribution in using chiplet-based systems as a sustainable alternatives to monolithic designs.  In addition to generating the plot, the script also prints the underlying raw data within the plot. 

\vspace{-1mm}

%%%%%%%%%%%%%%%%%%%%%%%%%%%%%%%%%%%%%%%%%%%%%%%%%%%%%%%%%%%%%%%%%%%%%
\subsection{Experiment customization}
\noindent
The experiment can be customized based on the input parameters that are provided to ECO-CHIP, \texttt{architecture.json} can customize the system architecture stating the chiplet sizes and types, along with package type that is used, in-depth packaging parameters can be customized under \texttt{packageC.json}. Parameters under \texttt{design.json} can be customized to explore more on the design CFP. Customizing \texttt{node\_list.txt} can help in exploring across different nodes. All the details for each of the input parameter files are explained under \texttt{ECO-CHIP/README.md}. 

%%%%%%%%%%%%%%%%%%%%%%%%%%%%%%%%%%%%%%%%%%%

%%%%%%%%%%%%%%%%%%%%%%%%%%%%%%%%%%%%%%%%%%%%%%%%%%%%%%%%%%%%%%%%%%%%%
\subsection{Methodology}

Submission, reviewing, and badging methodology:

\begin{itemize}
  \item \url{https://www.acm.org/publications/policies/artifact-review-badging}
  \item \url{http://cTuning.org/ae/submission-20201122.html}
  \item \url{http://cTuning.org/ae/reviewing-20201122.html}
\end{itemize}

\end{document}